\documentclass[prb,twocolumn]{revtex4}%

\usepackage{graphicx}
\usepackage{amsmath}
\usepackage{ulem}
\usepackage{color}

\newcommand{\be}{\begin{equation}}
\newcommand{\ee}{\end{equation}}
\newcommand{\bea}{\begin{eqnarray*}}
\newcommand{\eea}{\end{eqnarray*}}
\newcommand{\bean}{\begin{eqnarray}}
\newcommand{\eean}{\end{eqnarray}}

\begin{document}

\draft
\title{\bf Effects of Coulomb blockade on the charge transport through the
topological states of finite armchair graphene nanoribbons and
heterostructures}

\author{David M T Kuo}

\address{Department of Electrical Engineering and Department of Physics, National Central
University, Chungli, 320 Taiwan, China}

\date{\today}

\begin{abstract}
In this study, we investigate the charge transport properties of
semiconducting armchair graphene nanoribbons (AGNRs) and
heterostructures through their topological states (TSs), with a
specific focus on the Coulomb blockade region. Our approach
employs a two-site Hubbard model that takes into account both
intra- and inter-site Coulomb interactions. Using this model, we
calculate the electron thermoelectric coefficients and tunneling
currents of serially coupled TSs (SCTSs). In the linear response
regime, we analyze the electrical conductance ($G_e$), Seebeck
coefficient ($S$), and electron thermal conductance ($\kappa_e$)
of finite AGNRs. Our results reveal that at low temperatures, the
Seebeck coefficient is more sensitive to many-body spectra than
the electrical conductance. Furthermore, we observe that the
optimized $S$ at high temperature is less sensitive to electron
Coulomb interactions than $G_e$ and $\kappa_e$. In the nonlinear
response regime, we observe a tunneling current with negative
differential conductance through the SCTSs of finite AGNRs. This
current is generated by electron inter-site Coulomb interactions
rather than intra-site Coulomb interactions. Additionally, we
observe current rectification behavior in asymmetrical junction
systems of SCTSs of AGNRs. Notably, we also uncover the remarkable
current rectification behavior of SCTSs of 9-7-9 AGNR
heterostructure in the Pauli spin blockade configuration. Overall,
our study provides valuable insights into the charge transport
properties of TSs in finite AGNRs and heterostructures. We
emphasize the importance of considering electron-electron
interactions in understanding the behavior of these materials.
\end{abstract}

\maketitle

\section{Introduction}
The study of two-dimensional (2D) materials has gained significant
attention since the discovery of graphene in 2004
[\onlinecite{NovoselovK}-\onlinecite{NovoselovKS}]. Although
graphene has limited applications in optical, semiconductor, and
thermoelectric devices due to its gapless semi-metal nature, other
2D materials, such as $MoS_2$ and $WSe_2$, have shown potential
for use in transistors and optoelectronics due to their direct
band gaps. Low electron mobility and high contact resistance are
two of the main challenges that need to be overcome to improve the
performance of two-dimensional (2D) electronic devices, such as
those made of $MoS_2$ and $WSe_2$[\onlinecite{ShenPC}]. In recent
years, researchers have explored the use of heterostructures made
of 2D materials to improve the performance of quantum devices with
ballistic transport[\onlinecite{Iannaccone}].

To implement high power output quantum devices, it is crucial to
reduce the high contact resistance of 2D materials caused by
contacted metallic electrodes [\onlinecite{Iannaccone}]. On the
other hand, certain low-power output quantum devices, such as
single-electron transistors [\onlinecite{GuoLJ}], single-photon
emitters [\onlinecite{Michler}-\onlinecite{Gustavsson}],
spin-current conversion devices [\onlinecite{Ono}],
single-quantum-dot heat engines [\onlinecite{Josefsson}], and
solid-state quantum bits [\onlinecite{Cai}--\onlinecite{SunQ}],
require high contact resistances. Hence, 2D material
nanostructures with small dielectric constants may have promising
applications in low-power quantum devices. The exploration of
electronic structures with deep energy levels that are
well-separated from the band states in 2D material nanostructures
is crucial for their development since these quantum devices are
operated on the basis of few discrete states that are
well-separated from other continuous states to reduce charge
transport density.  A variety of graphene nanoribbons (GNRs) have
been extensively studied, including armchair GNRs
(AGNRs)[\onlinecite{SonYW}], AGNR heterostructures
[\onlinecite{CaoT}], cove-edged zigzag GNRs [\onlinecite{LinKS}],
and chevron GNRs[\onlinecite{JiangJ}]. These GNRs have
demonstrated the ability to exhibit diverse electronic topological
states (TSs) based on their width, edge shape, and end
terminations. The controllable manipulation of topological
invariants in materials is a highly pursued research area. For
instance, the utilization of electric fields and lattice strains
has been explored to modulate TSs
[\onlinecite{ZhaoF}-\onlinecite{Tepliakov}].

Significant progress has been made in the fabrication of graphene
nanoribbons (GNRs) and heterostructures using the bottom-up
synthesis technique [\onlinecite{Cai}--\onlinecite{SunQ}]. GNRs
can be classified into two categories: armchair GNRs (AGNRs) and
zigzag GNRs (ZGNRs). The topological states (TSs) of the end
zigzag edges of semiconducting AGNRs and interface states of AGNR
heterostructures are well-separated from the conduction and
valence subbands [\onlinecite{DRizzo},\onlinecite{DJRizzo}].
Therefore, these TSs may have potential for realizing low power
devices operating at room temperature. Due to graphene's small
dielectric constant, electron Coulomb interactions are expected to
be significant in the TSs of finite AGNRs and heterostructures
[\onlinecite{Golor}]. Investigating the effects of electron
Coulomb interactions on low-power devices made of GNRs is
desirable [\onlinecite{Sols}--\onlinecite{TongC}]. However, to
date, there has been a lack of theoretical and experimental
analysis concerning the Coulomb blockade effect in charge
transport through serially coupled TSs (SCTSs) of AGNRs and AGNR
heterostructures [\onlinecite{SonY}-\onlinecite{Mangnus}]. While
various systems involving GNRs have exhibited topological phases
[\onlinecite{SonYW}-\onlinecite{Tepliakov}], the utilization of
AGNRs and AGNR heterostructures holds particular advantages for
the realization of low-power quantum devices and circuits. This
advantage arises from their symmetrical end-edge structures, which
can be easily connected to line-contacted electrodes using current
bottom-up synthesis techniques [\onlinecite{Mangnus}].

This study aims to investigate the charge transport mechanisms of
SCTSs in two distinct systems: the end zigzag edge states of
finite AGNRs and the topologically protected interface states of
AGNR heterostructures coupled to leads, as shown in Figure 1(a)
and 1(b), respectively. A two-site model is employed to accurately
represent the electrical conductance spectra resulting from charge
transport through their SCTSs. A two-site Hubbard model with
intra- and inter-site Coulomb interactions and Green's function
techniques are used to reveal the effects of Coulomb blockade on
the charge transport of SCTS. In the linear response regime, we
calculate the electrical conductance ($G_e$), Seebeck coefficient
($S$), and electron thermal conductance ($\kappa_e$) of AGNRs'
SCTSs. The Seebeck coefficient shows greater sensitivity to
many-body spectra than electrical conductance at low temperatures.
At high temperatures, electron Coulomb interactions do not affect
the optimized values of $S$. In the nonlinear response regime, we
observe a tunneling current exhibiting negative differential
conductance (NDC). NDC arises due to inter-site electron Coulomb
interactions in the absence of bias-dependent orbital offset.
Moreover, in the asymmetrical tunneling junction systems of finite
AGNRs, we observe current rectification behavior due to inter-site
electron Coulomb interactions. Since the wave functions of TSs are
far away from the contacted electrodes, 9-7-9 AGNR
heterostructures can be readily set up in the Pauli spin blockade
configuration. As a result, tunneling current in a certain applied
bias direction is strongly suppressed due to SCTSs that are highly
occupied by two electron triplet states. This current
rectification feature in the PSB configuration is very useful for
spin-current conversion applications [\onlinecite{Ono}].

\begin{figure}[h]
\centering
\includegraphics[trim=2.5cm 0cm 2.5cm 0cm,clip,angle=0,scale=0.3]{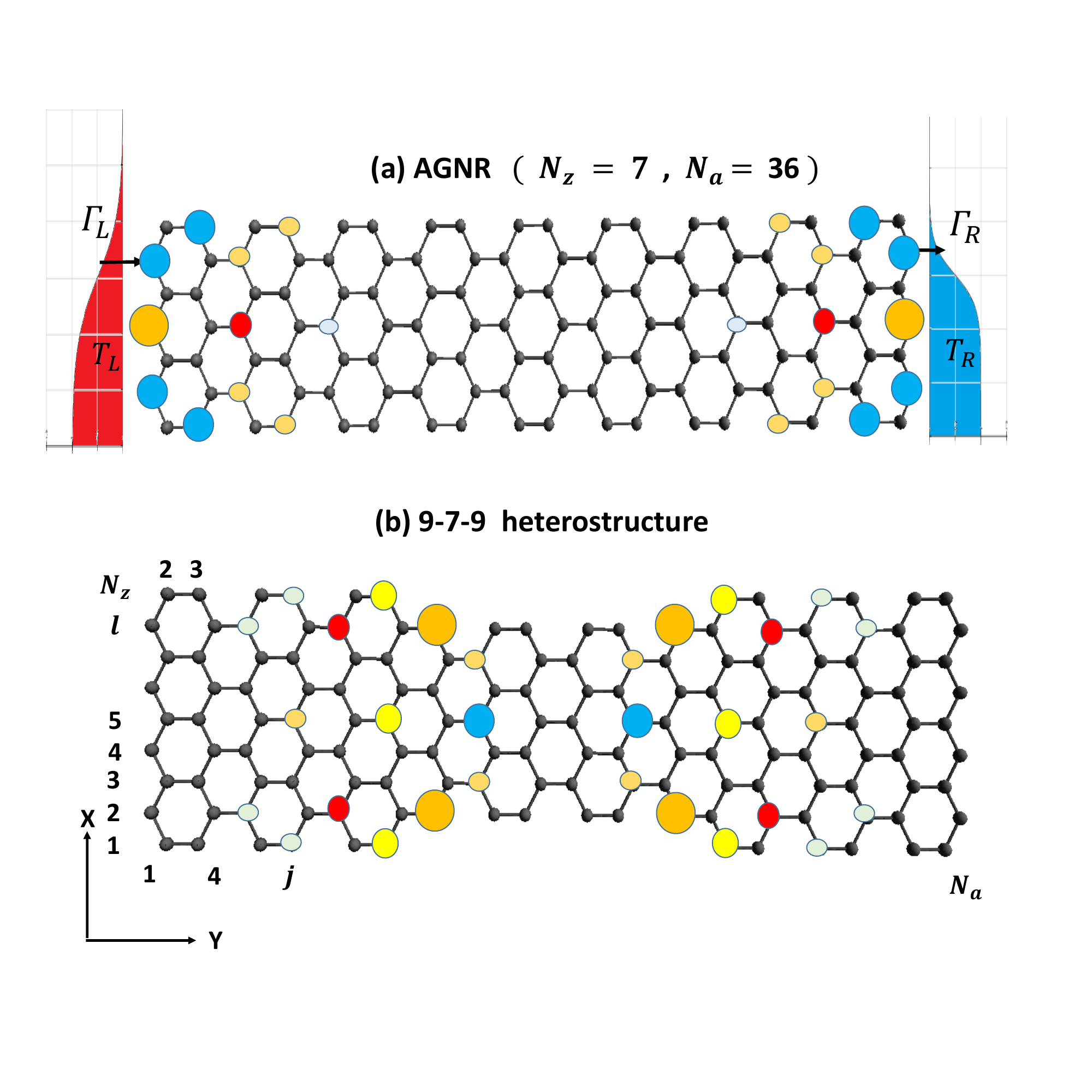}
\caption{(a) Schematic diagram of an AGNR connected to electrodes.
The tunneling rate of electrons between the left (right) electrode
and the leftmost (rightmost) atoms of the AGNR is denoted by
$\Gamma_{L}$ ($\Gamma_R$), respectively. The temperature of the
left ($L$) and right ($R$) electrodes is represented by $T_L$ and
$T_R$, respectively. The charge density of the localized zigzag
edge state with energy $\varepsilon = 9.087$ meV is shown for an
AGNR with size characterized by $(N_z, N_a) = (7, 36)$. (b)
Schematic lattice structure of AGNR heterostructures formed by
three segments. Each segment is characterized by $(N_z = 9(7), N_a
= 12)$. Note that $N_a = 12 $ correspond to three unit cells
(u.c.). The charge density of the topological state with energy
$\varepsilon = 0.12156$ eV is plotted for a 9/7/9 AGNR
heterostructure. The radius of the circle represents the intensity
of the charge density.}
\end{figure}

\section{Calculation method}
To model the transport properties of finite AGNRs and
heterostructures connected to the electrodes shown in Fig. 1, it
is a good approximation to employ a tight-binding model with one
$p_z$ orbital per atomic site to describe the electronic
states[\onlinecite{Nakada}-\onlinecite{Wakabayashi2}]. The
Hamiltonian of the nano-junction system depicted in Fig.~1,
including two different AGNR structures, can be written as
$H=H_0+H_{AGNR}$ [\onlinecite{Haug}], where
\begin{small}
\begin{eqnarray}
H_0& = &\sum_{k} \epsilon_k a^{\dagger}_{k}a_{k}+
\sum_{k} \epsilon_k b^{\dagger}_{k}b_{k}\\
\nonumber &+&\sum_{\ell}\sum_{k}
V^L_{k,\ell,j}d^{\dagger}_{\ell,j}a_{k}
+\sum_{\ell}\sum_{k}V^R_{k,\ell,j}d^{\dagger}_{\ell,j}b_{k} + H.c.
\end{eqnarray}
\end{small}
The first two terms of Eq.~(1) describe the free electrons in the
left ($L$) and right ($R$) electrodes. $a^{\dagger}_{k}$
($b^{\dagger}_{k}$) creates  an electron with wave number $k$ and
energy $\epsilon_k$ in the left (right) electrode.
$V^L_{k,\ell,j=1}$ ($V^R_{k,\ell,j=N_z(N_a)}$) describes the
coupling between the left (right) lead with its adjacent atom in
the $\ell$-th row. The Hamiltonian for AGNRs can be expressed as:

\begin{small}
\begin{eqnarray}
H_{AGNR} &= &\sum_{\ell,j} E_{\ell,j}~d^{\dagger}_{\ell,j}d_{\ell,j}\\
\nonumber&-& \sum_{\ell,j}\sum_{\ell',j'} t_{(\ell,j),(\ell', j')}
d^{\dagger}_{\ell,j} d_{\ell',j'} + h.c,
\end{eqnarray}
\end{small}

where $E_{\ell,j}$ is the on-site energy for the $p_z$ orbital in
the ${\ell}$-th row and $j$-th column. Here, the spin-orbit
interaction is neglected. $d^{\dagger}_{\ell,j} (d_{\ell,j})$
creates (destroys) one electron at the atom site labeled by (
$\ell$ , $j$ ) where $\ell$ and $j$, respectively are the row and
column indices as illustrated in Fig.~1. $t_{(\ell,j),(\ell',
j')}$ describes the electron hopping energy from site ( $\ell'$ ,
$j'$ ) to site ( $\ell$ , $j$ ). The tight-binding parameters used
for AGNRs is $E_{\ell,j} = 0$ for the on-site energy and
$t_{(\ell,j),(\ell',j')} = t_{pp\pi} = 2.7$ eV for the nearest
neighbor hopping strength.

To study the transport properties of an AGNR junction connected to
electrodes, it is convenient to use the Keldysh Green function
technique [\onlinecite{Haug}]. In the linear response regime, the
electrical conductance ($G_e$), Seebeck coefficient ($S$), and
electron thermal conductance ($\kappa_e$) are given by
$G_e=e^2{\cal L}_{0}$, $S=-{\cal L}_{1}/(eT{\cal L}_{0})$, and
$\kappa_e=\frac{1}{T}({\cal L}_2-{\cal L}^2_1/{\cal L}_0)$,
respectively. Here, ${\cal L}_n$ ($n = 0,1,2$) is defined as

\begin{equation}
\label{eq2} {\cal L}_n=\frac{2}{h}\int d\varepsilon~ {\cal
T}_{LR}(\varepsilon)(\varepsilon-\mu)^n\frac{\partial
f(\varepsilon)}{\partial \mu}.
\end{equation}

The Fermi distribution function of electrodes at equilibrium
temperature $T$ and chemical potential $\mu$ is given by
$f(\varepsilon)=1/(exp^{(\varepsilon-\mu)/k_BT}+1)$. The
transmission coefficient ${\cal T}_{LR}(\varepsilon)$, as shown in
Equation (\ref{eq2}), is a critical factor in electron transport
between the left ($L$) and right ($R$) electrodes. The numerical
code can be used to calculate ${\cal T}_{LR}(\varepsilon)$, which
is given by ${\cal
T}_{LR}(\varepsilon)=4Tr[\Gamma_L(\varepsilon)G^r(\varepsilon)\Gamma_R(\varepsilon)G^a(\varepsilon)]$.
Here, $\Gamma_L(\varepsilon)$ and $\Gamma_R(\varepsilon)$
represent the tunneling rate (in energy units) at the left and
right leads, respectively. Additionally, $G^{r}(\varepsilon)$ and
$G^{a}(\varepsilon)$ correspond to the retarded and advanced Green
functions of the AGNR, respectively
[\onlinecite{Kuo1}-\onlinecite{Kuo3}].

\section{Results and discussion}
\subsection{Charge transport through a Finite Armchair Graphene Nanoribbon (AGNR)}
The electronic behavior of AGNRs is determined by their widths,
which follow the rule $N_z = 3m + 2$, $N_z = 3m + 1$, and $N_z =
3m$, where $m$ is an integer. AGNRs exhibit semiconducting
behavior for $N_z = 3m + 1$ and $N_z = 3m$, resulting in
semiconducting phases for widths such as $N_z = 7$, $N_z = 9$, and
$N_z = 13$ corresponding band gaps of $1.26$ eV, $0.948$ eV, and
$0.714$ eV, respectively, in the absence of electron Coulomb
interactions [\onlinecite{Wakabayashi},\onlinecite{Wakabayashi2}].
Notably, AGNRs with $N_z = 3m + 2$ maintain their semiconducting
phases, as determined by first-principle calculations in
references [\onlinecite{SonYW},\onlinecite{Tepliakov}]. However,
the one-band tight binding model still captures the main physics
for the cases of $N_z = 3m$ and $N_z = 3m+1 $, which are the
primary focus of this article. The band gap of AGNRs decreases as
their width increases. To investigate charge transport through
AGNRs with line contacted electrodes, as depicted in Fig. 1, we
present the calculated electron conductance ($G_e$) as a function
of $\mu$ at zero temperature for various $N_z$ values of finite
AGNRs with $N_a = 40$ in Fig. 2. The magnitude of $G_e$ is given
in units of the quantum conductance of
$G_0=\frac{2e^2}{h}=1/(12.9K\Omega)=77.5\mu S$. For the case of
$N_z = 7$ in Fig. 2(a), we observe a peculiar peak labeled by
$\Sigma_0$ at the charge neutrality point (CNP) ($\mu=0$). In
contrast, the spectra around the CNP are vanishingly small for
$N_z= 9$, as seen in Fig. 2(b). The uniform peak separation in
Fig. 2(c) corresponds to the linear dispersion of AGNRs with a
metallic phase as $N_z = 11$. In particular, $\Sigma_0$ is split
into two peaks around the CNP for $N_z = 13$ in Fig. 2(d). The
electrical conductance near the CNP is attributed to electron
transport through the end zigzag edges of finite AGNRs. The wave
functions of the left and right zigzag edges of AGNRs decay along
the armchair directions, and their overlaps are vanishingly small
in infinitely long AGNRs. When $N_a$ is finite, the zero mode
energy levels are lifted due to the coupling between these two
localized wave functions which can be regarded as a SCTS. The
magnitude of the lifted energy level can be determined by the
electron hopping strength $t_{LR}$ between the left and right
zigzag edges [\onlinecite{DJRizzo}]. We have $|t_{LR}|= 5.323$
meV, $0.1$ meV, and $39.67$ meV for $N_z = 7$, $N_z = 9$, and $N_z
= 13$, respectively. Since $|t_{LR}|/\Gamma_t \ll 1$, the peak of
$\Sigma_0$ is degraded in the case of $N_z = 9$. The results of
Fig. 2 indicate that $t_{LR}$ is significantly affected by the
AGNR width, which is either $N_z=3m + 1$ or $N_z= 3m$. In appendix
A, we discuss the electronic structures of 9-7 AGNR superlattices
and charge transport through the SCTSs of 9-7-9 AGNR
heterostructures, which were proposed as useful quantum bits
[\onlinecite{DJRizzo}].

\begin{figure}[h]
\centering
\includegraphics[angle=0,scale=0.3]{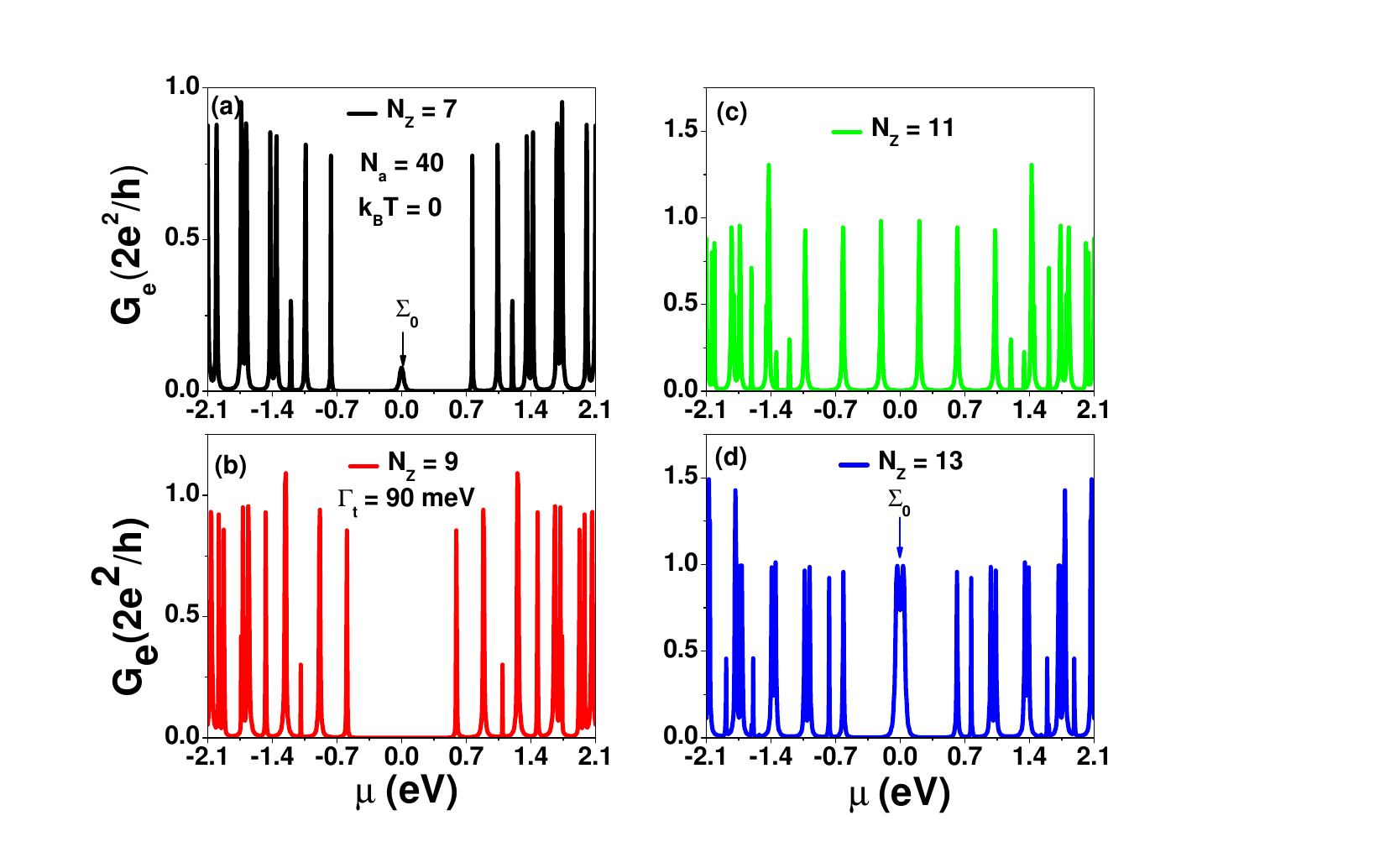}
\caption{ Electrical conductance ($G_e$) as a function of chemical
potential ($\mu$) for AGNRs with a fixed armchair direction width
of $N_a = 40$ ($L_a = 4.12$ nm) and various widths ( $N_z$ ) at a
tunneling rate of $\Gamma_L = \Gamma_R = \Gamma_t = 90$~meV and
zero temperature ($T = 0$~K). (a) $N_z = 7$, (b) $N_z = 9$, (c)
$N_z = 11$, and (d) $N_z = 11$. }
\end{figure}

To better understand the transport behavior of the SCTSs, Figure
3(a) shows the calculated electrical conductance ($G_e$) as a
function of $\mu$ for various $N_a$ at $\Gamma_t= 90$~meV, $T = 0
K$ and $N_z = 13$. As $N_a$ increases for a fixed $N_z = 13$,
$t_{LR}$ decreases, with values of $39.67$~meV, $22.29$~meV,
$12.59$~meV, and $7.13$ meV for $N_a = 40$, $N_a = 48$, $N_a =
56$, and $N_a = 64$, respectively. As seen in Fig. 3(a),
$\varepsilon_{HO}$ and $\varepsilon_{LU}$, which are the highest
occupied molecular orbital (HOMO) and the lowest unoccupied
molecular orbital (LUMO), respectively, can be resolved only for
$N_a = 40$. For $N_a = 64$, the magnitude of electrical
conductance is smaller than 0.4 $G_0$. To clarify the effect of
contacted electrodes on the transport of the SCTS, we plot the
calculated $G_e$ for various $\Gamma_t$ values for $N_a = 64$ and
$N_z = 13$ in Fig. 3(b). As $\Gamma_t$ decreases, the electrical
conductance is highly enhanced, with $G_e$ almost reaching one
quantum conductance $G_0$ at $\Gamma_t = 27 $~meV.

\begin{figure}[h]
\centering
\includegraphics[angle=0,scale=0.3]{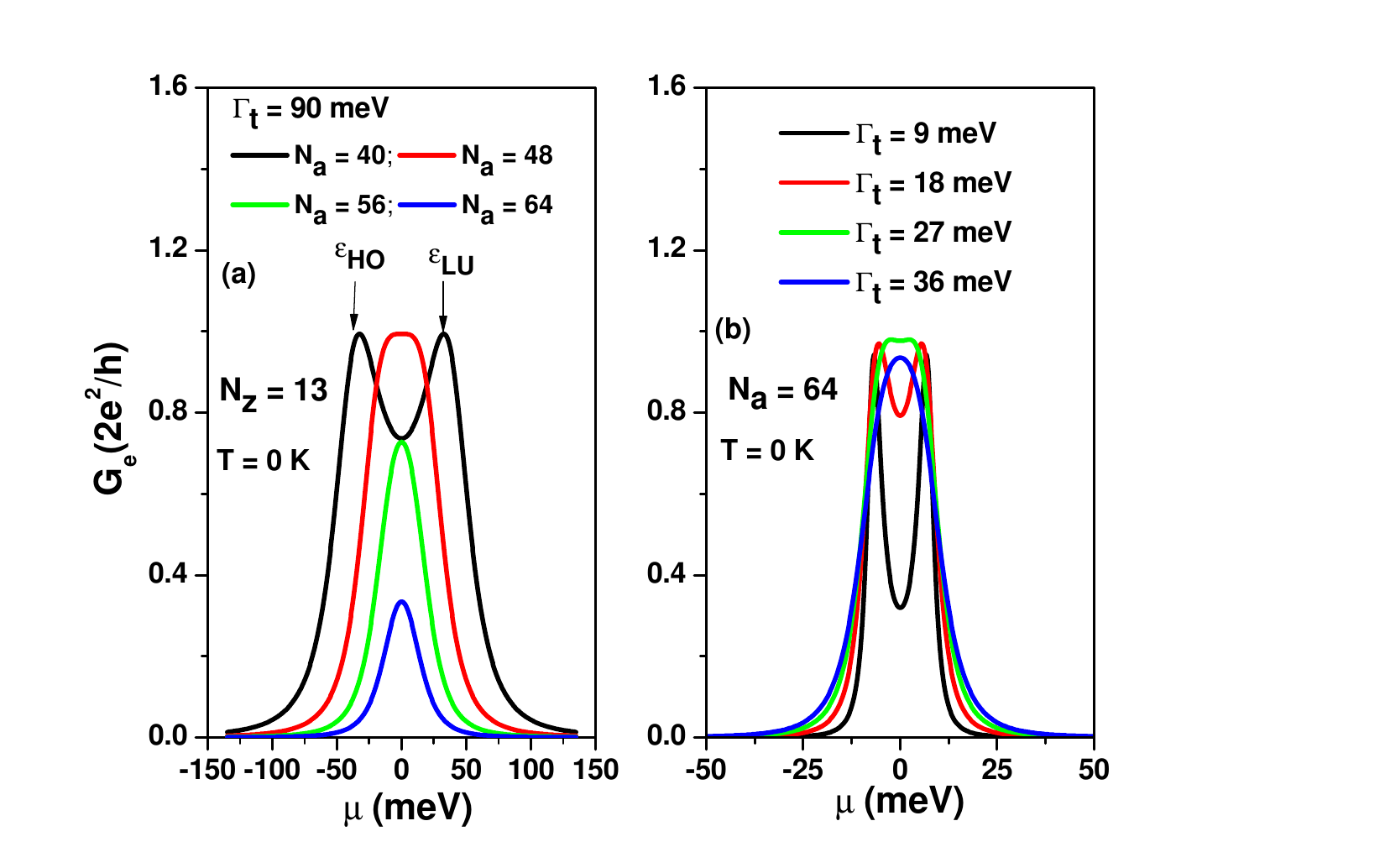}
\caption{ (a) Electrical conductance $G_e$ as a function of
chemical potential $\mu$ for AGNRs of different lengths, while
keeping the tunneling rate fixed at $\Gamma_t=90$~meV and the
temperature at $T = 0$~K. (b) Electrical conductance $G_e$ as a
function of chemical potential $\mu$ for AGNRs with $N_a = 64$
($L_a= 6.674$~nm), while varying the tunneling rate $\Gamma_t$ at
$T = 0 K$.}
\end{figure}

Up to now, it remains unclear about the thermoelectric properties
of SCTSs [\onlinecite{DRizzo}-\onlinecite{SunQ}]. In Fig. 4, we
present the calculated electrical conductance $G_e$, Seebeck
coefficient $S$, power factor $PF=S^2G_e$, and thermoelectric
figure of merit $ZT$ as functions of chemical potential $\mu$ for
various temperatures for AGNRs with $N_z = 13$ and $N_a = 64$. The
spectra of $G_e$ exhibit a symmetrical function of $\mu$, while
$S$ displays a bipolar behavior with respect to $\mu$. At $T = 48
K$, the maximum $S_{h(e),max} = \pm 2.753$. Notably, $S$ can be
well described by $S = \mu/T$ at room temperature ($T = 324 K$),
indicating that the thermoelectric properties of zigzag edge
states of AGNRs with $N_z = 13$ and $N_a = 64$ resemble those of a
single localized state at room temperature
[\onlinecite{MahanGD}-\onlinecite{XuY}]. In Fig. 4(c), the maximum
power factors ($PF_{h(e),max}$) are $0.76$ and $0.339$ for $T = 48
K$ and $T = 324 K$, respectively. At $T = 324 K$, $PF_{h(e),max}$
occurs at $\mu = \pm 66.68$ meV, with the ratio of $\mu/k_BT = \pm
2.46$. In ref. [\onlinecite{MahanGD}], the authors predicted that
the optimized $PF$ of a single energy level without energy level
broadening satisfies $\mu/k_BT = \pm 2.4$. In Fig. 4(d), we
calculated $ZT$ using the equation $ZT = S^2G_e
T/(\kappa_e+\kappa_{ph})$. Here, we used a phonon thermal
conductance value of $\kappa_{ph} = F_s*\frac{\pi^2 k^2_B T}{3h}$,
where we assumed a scattering factor of $F_s = 0.1$. For the
remainder of this article, we will use the same value of
$\kappa_{ph} = F_s*\frac{\pi^2 k^2_B T}{3h}$ with $F_s=0.1$. In a
previous study [\onlinecite{XuY}], the authors proposed
introducing defects in GNRs to increase phonon scattering while
maintaining the electron transmission coefficient of topological
states. More recently, we demonstrated [\onlinecite{Kuo1}] that
zigzag edge states in AGNRs with vacancies remained robust. The
maximum $ZT_{h(e),max} = 1.966$ at $T = 324 K$ occurs at $\mu= \pm
68.3$ meV. Here and henceforth, $S$ is expressed in units of
$k_B/e = 86.25~\mu V/K$, and $PF$ is in units of $2k^2_B/h =
0.575~pW/K^2$. The results of Fig. 4 provide us with
thermoelectric coefficients of the SCTS in the absence of electron
Coulomb interactions.

\begin{figure}[h]
\centering
\includegraphics[angle=0,scale=0.3]{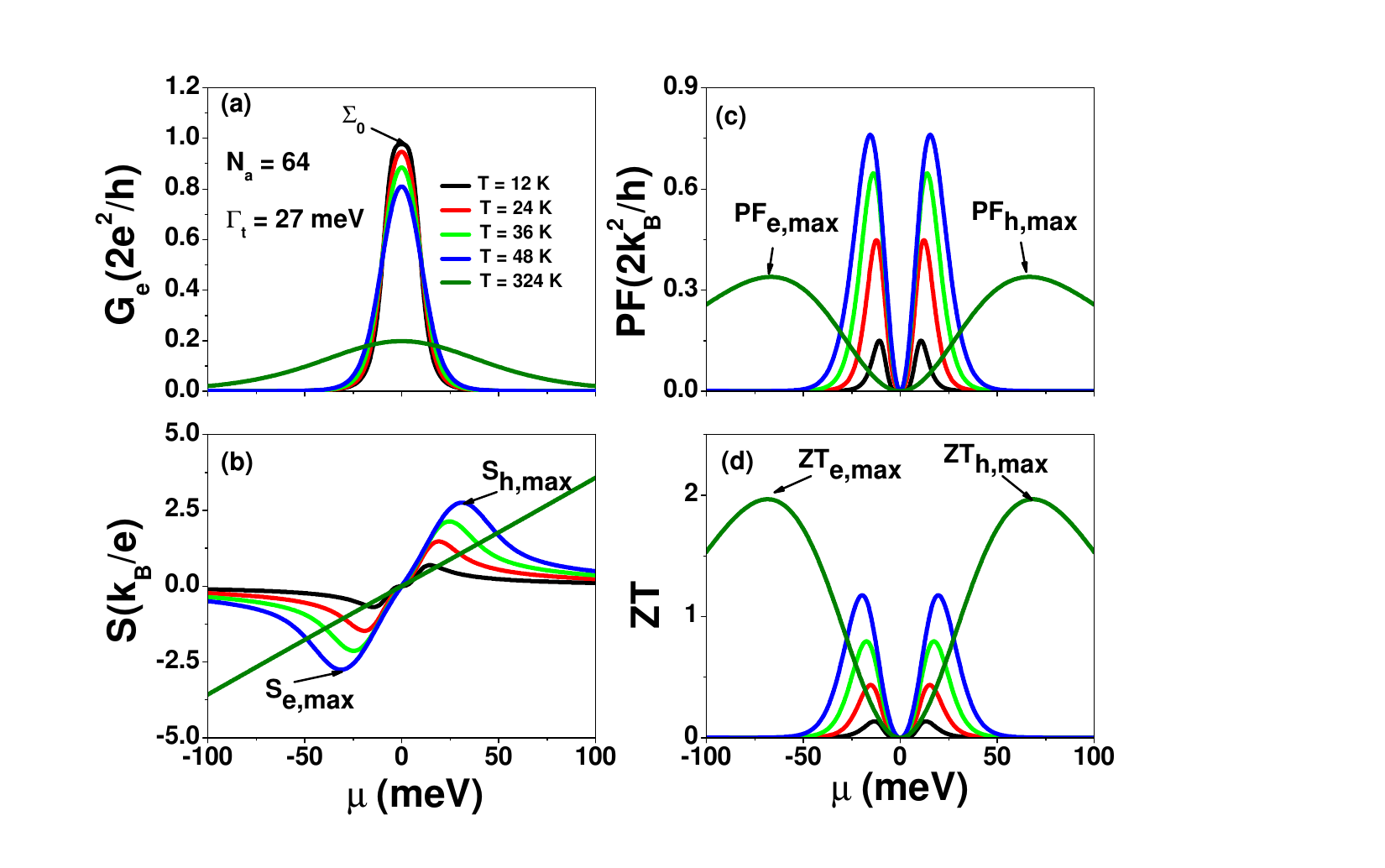}
\caption{(a) Electrical conductance $G_e$, (b) Seebeck coefficient
$S$, (c) power factor $PF=S^2G_e$, and (d) figure of merit $ZT =
S^2G_eT/(\kappa_e+\kappa_{ph})$ as functions of  $\mu$ for AGNRs
with $N_z = 13$, $N_a = 64$, and $\Gamma_t = 27$~meV at various
temperatures.}
\end{figure}

\subsection{Effective Hamiltonian and Formula for Tunneling Current}
It is challenging to calculate tunneling current through SCTSs in
the Coulomb blockade region using either first-principles methods
[\onlinecite{SonYW}] or tight-binding models [\onlinecite{Kuo2}].
Hence, we introduce a two-site Hubbard model to clarify the
thermoelectric coefficients and tunneling current of SCTSs in the
Coulomb blockade region. The Hamiltonian of the two-site Hubbard
model is given by
\begin{small}
\begin{eqnarray}
& &H_{2-site}\\ \nonumber &=&
\sum_{j=L,R,\sigma}E_{j}c^{\dagger}_{j,\sigma}c_{j,\sigma} -
t_{LR}~(c^{\dagger}_{R,\sigma}c_{L,\sigma} +
c^{\dagger}_{L,\sigma}c_{R,\sigma})\\ \nonumber & + &
\sum_{j=L,R}U_j~n_{j,\sigma}n_{j,-\sigma} +
\frac{1}{2}\sum_{j\neq\ell,\sigma,\sigma'}U_{j,\ell}~n_{j,\sigma}n_{\ell,\sigma'},
\end{eqnarray}
\end{small}
where $E_j$ represents the spin-independent energy level of the
TSs, $U_j = U_{L(R)} = U_0$ and $U_{j,\ell} = U_{LR} = U_1$ denote
the intra-site and inter-site Coulomb interactions, respectively,
and $n_{j,\sigma} = c^{\dagger}_{j,\sigma}c_{j,\sigma}$. $U_0$ and
$U_1$ are calculated using
$\frac{1}{4\pi\epsilon_0}\sum_{i,j}|\Psi_{L(R)}(\textbf{r}_i)|^2|\Psi_{L(R)}(\textbf{r}_j)|^2\frac{1}{|\textbf{r}_i-\textbf{r}_j|}$
with the dielectric constant $\epsilon_0 = 4$, and $U_{cc} = 4$ eV
at $i = j$. $U_{cc}$ arises from the two-electron occupation in
each $p_z$ orbital. $\Psi_{L(R)}(\textbf{r}_i)$ is the wave
functions of TSs[\onlinecite{Golor}].

Based on the effective Hamiltonian given by Eq. (4), we derive the
tunneling current through the SCTSs coupled to the electrodes
after tedious algebra [\onlinecite{Kuo5}]. We obtain the
expression of tunneling current leaving from the left (right)
electrode as
\begin{eqnarray}
&&J_{L(R)}(V_{bias},T)\\ \nonumber &=&\frac{2e}{h}\int
{d\varepsilon}~ {\cal
T}^{2-site}_{LR(RL)}(\varepsilon)[f_L(\varepsilon)-f_R(\varepsilon)].
\end{eqnarray}
The Fermi distribution function for the $\alpha$-th electrode is
denoted by $f_{\alpha}(\varepsilon) =
1/(\exp[(\varepsilon-\mu_{\alpha})/k_BT]+1)$, where $\mu_{L(R)}$
is the chemical potential of the left (right) electrode with an
applied bias of $V_{bias}/2$ ($-V_{bias}/2$), such that
$\mu_{L(R)} = \mu \pm eV_{bias}/2$. The transmission coefficient
for charge transport through the SCTSs, denoted by ${\cal
T}^{2-site}_{LR}(\varepsilon)$, has a closed-form expression given
by

\begin{small}
\begin{eqnarray}
& &{\cal T}^{2-site}_{LR}(\epsilon)/(4t^2_{LR}\Gamma_{e,L}\Gamma_{e,R})=\frac{P_{1} }{|\epsilon_L\epsilon_R-t^2_{LR}|^2} \nonumber \\
&+& \frac{P_{2} }{|(\epsilon_L-U_{LR})(\epsilon_R-U_R)-t^2_{LR}|^2} \nonumber \\
&+& \frac{P_{3} }{|(\epsilon_L-U_{LR})(\epsilon_R-U_{LR})-t^2_{LR}|^2} \label{TF} \\
\nonumber &+&
\frac{P_{4} }{|(\epsilon_L-2U_{LR})(\epsilon_R-U_{LR}-U_R)-t^2_{LR}|^2}\\
\nonumber &+& \frac{P_{5} }{|(\epsilon_L-U_{L})(\epsilon_R-U_{LR})-t^2_{LR}|^2}\\
\nonumber &+& \frac{P_{6}
}{|(\epsilon_L-U_L-U_{LR})(\epsilon_R-U_R-U_{LR})-t^2_{LR}|^2}\\
\nonumber &+&
\frac{P_{7} }{|(\epsilon_L-U_L-U_{LR})(\epsilon_R-2U_{LR})-t^2_{LR}|^2}\\
\nonumber
 &+&
\frac{P_{8} }{|(\epsilon_L-U_L-2U_{LR})(\epsilon_R-U_R-2U_{LR})-t^2_{LR}|^2}, \\
\nonumber
\end{eqnarray}
\end{small}
where $\epsilon_L=\varepsilon-E_L+i\Gamma_{e,L}$ and
$\epsilon_R=\varepsilon-E_R+i\Gamma_{e,R}$. $\Gamma_{e,L(R)}$ is
an effective tunneling rate for the left (right) TS coupled to the
left (right) electrode. The transmission coefficient given by
Eq.(6) consists of eight terms, each corresponding to one of the
eight possible configurations of the SCTS that an electron with
spin $\sigma$ from the left electrode may encounter. The
probabilities of these configurations are as follows:

\begin{small}
\begin{eqnarray}
P_{1}&=&1-N_{L,\sigma}-N_{R,\sigma}-N_{R,-\sigma}+ \langle
n_{R,\sigma}n_{L,\sigma}\rangle \nonumber \\ &+&\langle
n_{R,-\sigma}n_{L,\sigma}\rangle+\langle
n_{R,-\sigma}n_{R,\sigma}\rangle-\langle
n_{R,-\sigma}n_{R,\sigma} n_{L,\sigma} \rangle \nonumber \\
P_{2}&=&N_{R,\sigma}-\langle n_{R,\sigma} n_{L,\sigma}\rangle
-\langle n_{R,-\sigma} n_{R,\sigma}\rangle \nonumber \\ &+&\langle
n_{R,-\sigma} n_{R,\sigma} n_{L,\sigma}\rangle \nonumber \\
P_{3}&=&N_{R,-\sigma}-\langle n_{R,-\sigma} n_{L,\sigma}\rangle
-\langle n_{R,-\sigma}n_{R,\sigma}\rangle \nonumber \\ &+&\langle
n_{R,-\sigma}n_{R,\sigma} n_{L,\sigma} \rangle \nonumber \\
P_{4}&=&\langle n_{R,-\sigma}n_{R,\sigma}\rangle-\langle
n_{R,-\sigma}n_{R,\sigma} n_{L,\sigma}\rangle \nonumber\\
P_{5}&=&N_{L,\sigma}- \langle n_{R,\sigma}n_{L,\sigma}\rangle
-\langle n_{R,-\sigma} n_{L,\sigma}\rangle \nonumber \\ &+&\langle
n_{R,-\sigma}n_{R,\sigma} n_{L,\sigma}\rangle \nonumber \\
P_{6}&=&\langle n_{R,\sigma} n_{L,\sigma}\rangle -\langle
n_{R,-\sigma}n_{R,\sigma} n_{L,\sigma}\rangle \nonumber \\
P_{7}&=&\langle n_{R,-\sigma}n_{L,\sigma}\rangle -\langle
n_{R,-\sigma}n_{R,\sigma} n_{L,\sigma}\rangle \nonumber \\
p_{8}&=&\langle n_{R,-\sigma}n_{R,\sigma}n_{L,\sigma} \rangle
\nonumber,
\end{eqnarray}
\end{small}
where the intra-site and inter-site two-particle correlation
functions are denoted by $\langle
n_{\ell,-\sigma}n_{\ell,\sigma}\rangle$ and $\langle
n_{\ell,\sigma}n_{j,\sigma}\rangle$ ($\langle
n_{\ell,-\sigma}n_{j,\sigma}\rangle$), respectively. $\langle
n_{\ell,-\sigma}n_{\ell,\sigma}n_{j,\sigma}\rangle$ is the
three-particle correlation function. These correlation functions
can be solved self-consistently, and it should be noted that
probability conservation is satisfied by $\sum_{m}P_{m}=1$.

The expression for ${\cal T}^{2-site}_{LR}(\varepsilon)$ goes
beyond not only mean field theory [\onlinecite{Mangnus}], but also
our previous work [\onlinecite{Kuo5}], where we only considered
one-particle occupation numbers and intra-site two-particle
correlation functions. Theoretical analysis has shown that
inter-site two-particle correlation functions play a significant
role when inter-site Coulomb interactions are large ($U_1 >
t_{LR}$) [\onlinecite{Kuo6}]. To obtain ${\cal
T}^{2-site}_{RL}(\varepsilon)$, we can simply exchange the indices
in Eq. (6). Based on Eq. (6), the expression for the electrical
conductance at zero temperature for a 2-site model without Coulomb
interactions is given by:

\begin{eqnarray}
& & G_e(\mu)/G_0\nonumber \\
&=&\frac{4\Gamma_{e,L}t^2_{LR}\Gamma_{e,R}}{|(\mu-E_L+i\Gamma_{e,L})(\mu-E_R+i\Gamma_{e,R})-t^2_{LR}|^2}.
\end{eqnarray}

Using Eq. (7), we calculated $G_e$ for different values of
$t_{LR}$ at $\Gamma_{e,L}=\Gamma_{e,R}=\Gamma_{e,t}=22.3$ meV in
Fig. 5(a). The considered $t_{LR}$ values correspond to $N_a =
40$, $N_a= 48$, $N_a= 56$, and $N_a= 64$. Similarly, we calculated
$G_e$ for various $\Gamma_{e,t}$ values at $t_{LR}=7.13$ meV in
Fig. 5(b). It is worth noting that the curves shown in Fig. 5(a)
and 5(b) are identical to the curves of Fig. 3(a) and 3(b),
respectively. These results in Fig. 5 illustrate that charge
transport through the SCTS can be well explained with a two-site
model that introduces an effective tunneling rate $\Gamma_{e,t}$.
In the following discussion, we will consider charge transport
through the SCTSs in the presence of electron Coulomb
interactions, both in the linear and nonlinear regimes.

\begin{figure}[h]
\centering
\includegraphics[angle=0,scale=0.3]{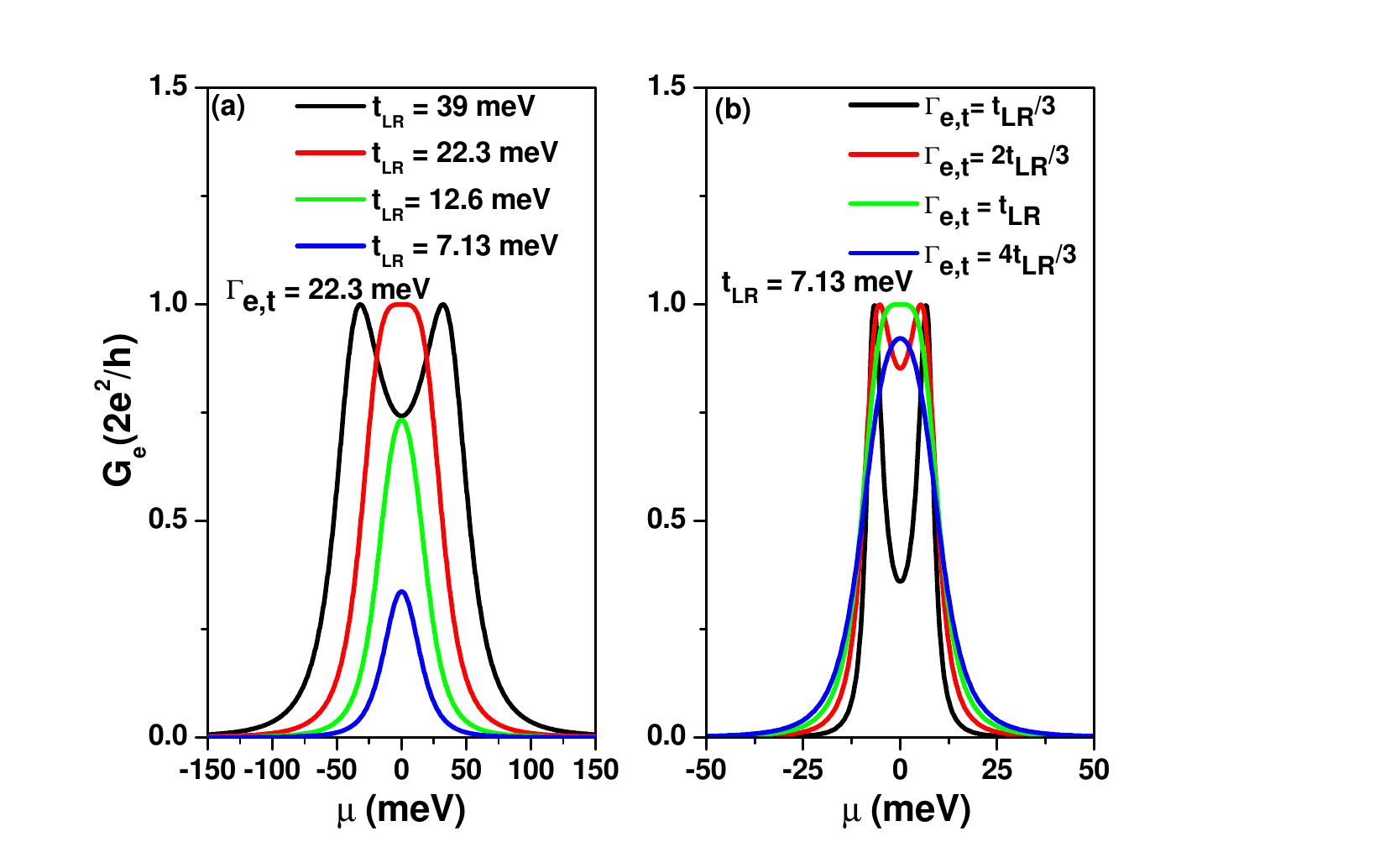}
\caption{(a) Electrical conductance as functions of $\mu$ for
various $t_{LR}$ values at zero temperature and
$\Gamma_{e,t}=22.3$~meV and (b) electrical conductance as
functions of $\mu$ for various $\Gamma_{e,t}$ values at zero
temperature and $t_{LR} = 7.13$~meV. The calculated lines is based
on Eq. (7).}
\end{figure}

\subsection{Effects of Coulomb Blockade on Charge Transport through the Zigzag Edges of AGNRs}
\subsubsection{Linear response regime}
In this subsection, we examine the effects of Coulomb blockade on
the charge transport through the end zigzag edges of AGNRs in the
linear and nonlinear response regimes. We utilize Eqs. (3) and (6)
to calculate the correlation functions (CF), electrical
conductance $G_e$, Seebeck coefficient $S$, and electron thermal
conductance $\kappa_e$ of the zigzag edge states as functions of
$\mu$ at $T = 1.2 K$ and $\Gamma_{e,t} = 1$ meV, and present the
results in Fig. 6.

Fig. 6(a) shows the electron single particle occupation number
$N_{j,\sigma}=\langle n_{j,\sigma} \rangle $, which displays four
major plateaus corresponding to the values of $1/4$, $1/2$, $3/4$,
and $1$ (or total electron number $N_T=\sum_{j,\sigma}
N_{j,\sigma}$ corresponding to 1, 2, 3, and 4). These plateaus
arise due to electron Coulomb interactions. In contrast to the
two-electron singlet state in each site $\langle n_{L,-\sigma}
n_{L,\sigma} \rangle $ and $\langle n_{R,-\sigma} n_{R,\sigma}
\rangle $, which occur at a large $\mu$ value near $197$ meV, the
inter-site triplet state $\langle n_{R,\sigma} n_{L,\sigma}\rangle
$ and the inter-site singlet state $\langle n_{R,-\sigma}
n_{L,\sigma} \rangle $ appear near $\mu \approx 42$ meV.

Fig. 6(b) presents the spectra of $G_e$, which can reveal electron
hopping strength ($t_{LR}=7.13$~meV), inter-site Coulomb
interactions ($U_1 = 42$~meV), and intra-site Coulomb interactions
($U_0= 155$~meV) due to low temperature and weak $\Gamma_{e,t}$.
The asymmetrical electrical conductances labeled by
$\varepsilon_{HO}$ and $\varepsilon_{LU}$ indicate that the
probability of $P_1$ depends on the $\mu$ of the electrodes. In
the Coulomb gap region ($N_{j,\sigma}=0.5$), there are two tiny
peaks labeled by $\varepsilon_{4(7)}$ and $\varepsilon_{2(5)}$
corresponding to $P_{4(7)}$ and $P_2(P_5)$ channels. Because they
are off-resonant channels as $E_L = E_R$, their electrical
conductances are quite small. In the Pauli spin blockade
configuration, they are in resonant channels, which will be
discussed later.

Fig. 6(c) shows the spectra of Seebeck coefficient $S$, which can
more clearly reveal the two tiny spectra of $G_e$
($\varepsilon_{4(7)}$ and $\varepsilon_{2(5)}$) because $S =
-\frac{\pi^2k^2_BT}{3e}\frac{1}{G_{e}(\mu)}\frac{\partial
G_e(\mu)}{\partial \mu}$ at low temperature. Fig. 6(d) displays
the spectra structure of electron thermal conductance $\kappa_e$,
which is quite similar to $G_e$ spectra at low temperature,
whereas they are quite different at high temperatures (see results
of Fig. 7 and Fig. 8). Note that $\kappa_e$ is in units of
$\kappa_0=0.62~nW/K$.

\begin{figure}[h]
\centering
\includegraphics[angle=0,scale=0.3]{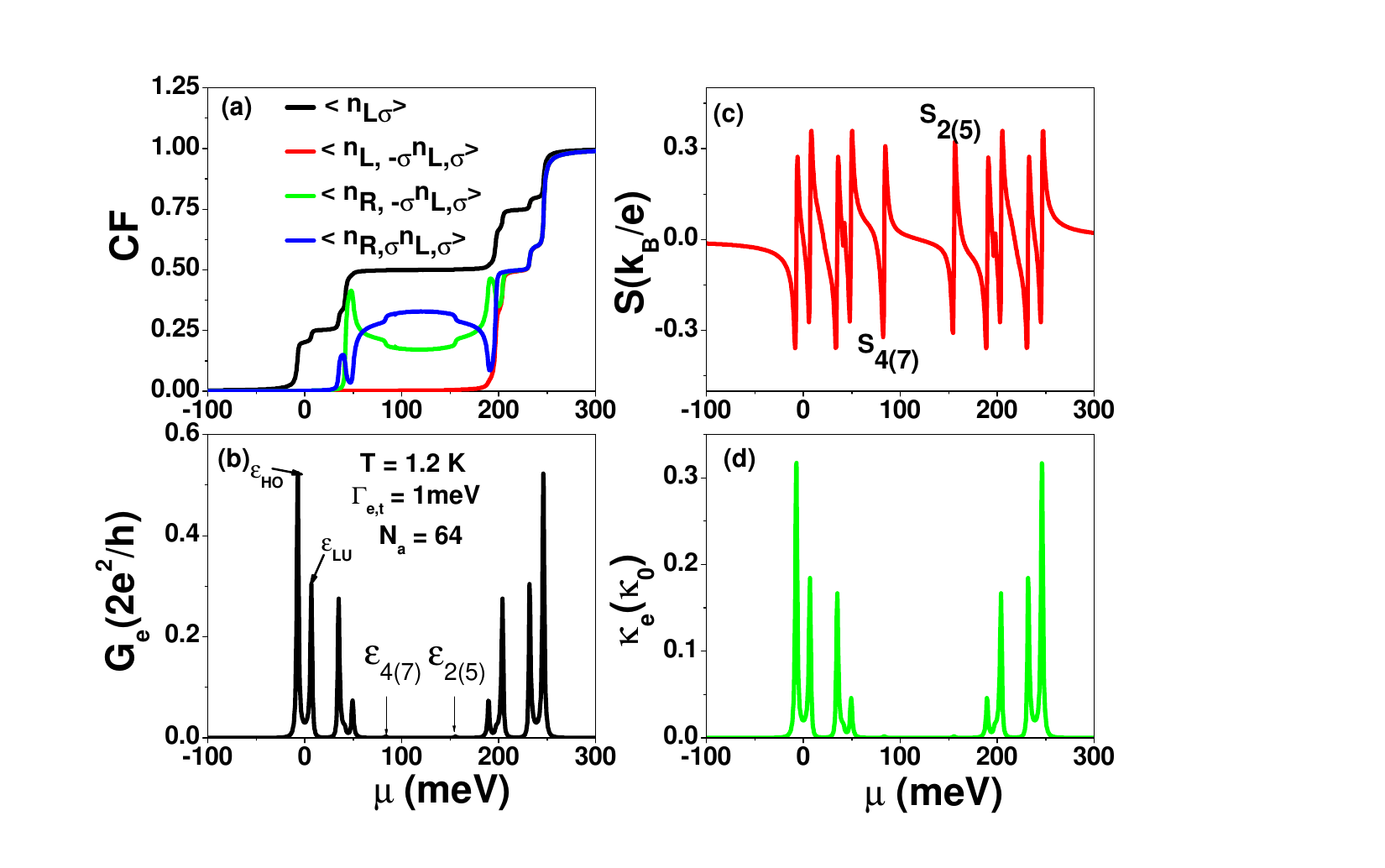}
\caption{(a) Correlation functions (CF), (b) electrical
conductance ($G_e$), (c) Seebeck coefficient $S$ and (d) electron
thermal conductance ($\kappa_e$) as functions of chemical
potential $\mu$ at $\Gamma_{e,t} = 1 $~meV, and  $T = 1.2 K$. We
have set $E_j = 0$, $U_0 = 155$ meV, $U_1 = 42$ meV and $t_{LR} =
7.13$ meV. These physical parameters correspond to AGNRs with $N_a
= 64$ and $N_z$= 13. }
\end{figure}

To clarify the contact properties in the presence of Coulomb
interactions, we then calculated several thermoelectric
quantities, including $G_e$, $S$, $\kappa_e$, $PF$,
$\rho=\kappa_e/(TG_e)$ and $ZT$, as functions of $\mu$ for various
$\Gamma_{e,t}$ at a temperature of $48 K$, and plotted the results
in Fig. 7. Consistent with the results in Fig. 5, the maximum
$G_e$ still occurred at the condition of $\Gamma_{e,t} = t_{LR}$,
which is not affected by the intra-site and inter-site Coulomb
interactions. At $T = 48 K$, the $\varepsilon_{HO}$ and
$\varepsilon_{LU}$ peaks are washed out in the spectra of $G_e$,
whereas the Coulomb gap between $\varepsilon_2$ and
$\varepsilon_3$ arising from intra-site Coulomb interactions
remains. Although the spectra of $S$ become more complex in the
presence of electron Coulomb interactions, it does not
significantly affect the maximum $S_{h(e),max}$. Comparing with
the results of $S_{h(e),max}$ at $T = 48$ shown in Fig. 4, the
maximum $S_{h(e),max}$ has a value of $\pm 2.684$ at
$\Gamma_{e,t}=t_{LR}$, which is almost the same as $S_{h(e),max}$
of Fig. 4. The spectra of $\kappa_e$ show two main structures,
each with three peaks, and an extra peak appear in the charge
blockade region (as seen in the $G_e$ spectra) as temperature
increases. The maximum $PF$ does not occur at $\Gamma_{e,t} =
t_{LR}$, as seen in Fig. 7(d). The Lorenz number of charge
transport through the zigzag edges does not satisfy the
Wiedemann-Franz law ${\cal L}_z=\kappa_e/(T
G_e)=\frac{\pi^2k^2_B}{3e^2}$ in Fig. 7(e), indicating that
$\kappa_e$ and $G_e$ may not be relevant thermoelectric quantities
in discrete energy level systems. Fig. 7(f) shows that the maximum
$ZT$ values prefer smaller $\Gamma_{e,t}$ and
$\rho=\kappa_e/(TG_e) < {\cal L}_z$.


\begin{figure}[h]
\centering
\includegraphics[angle=0,scale=0.3]{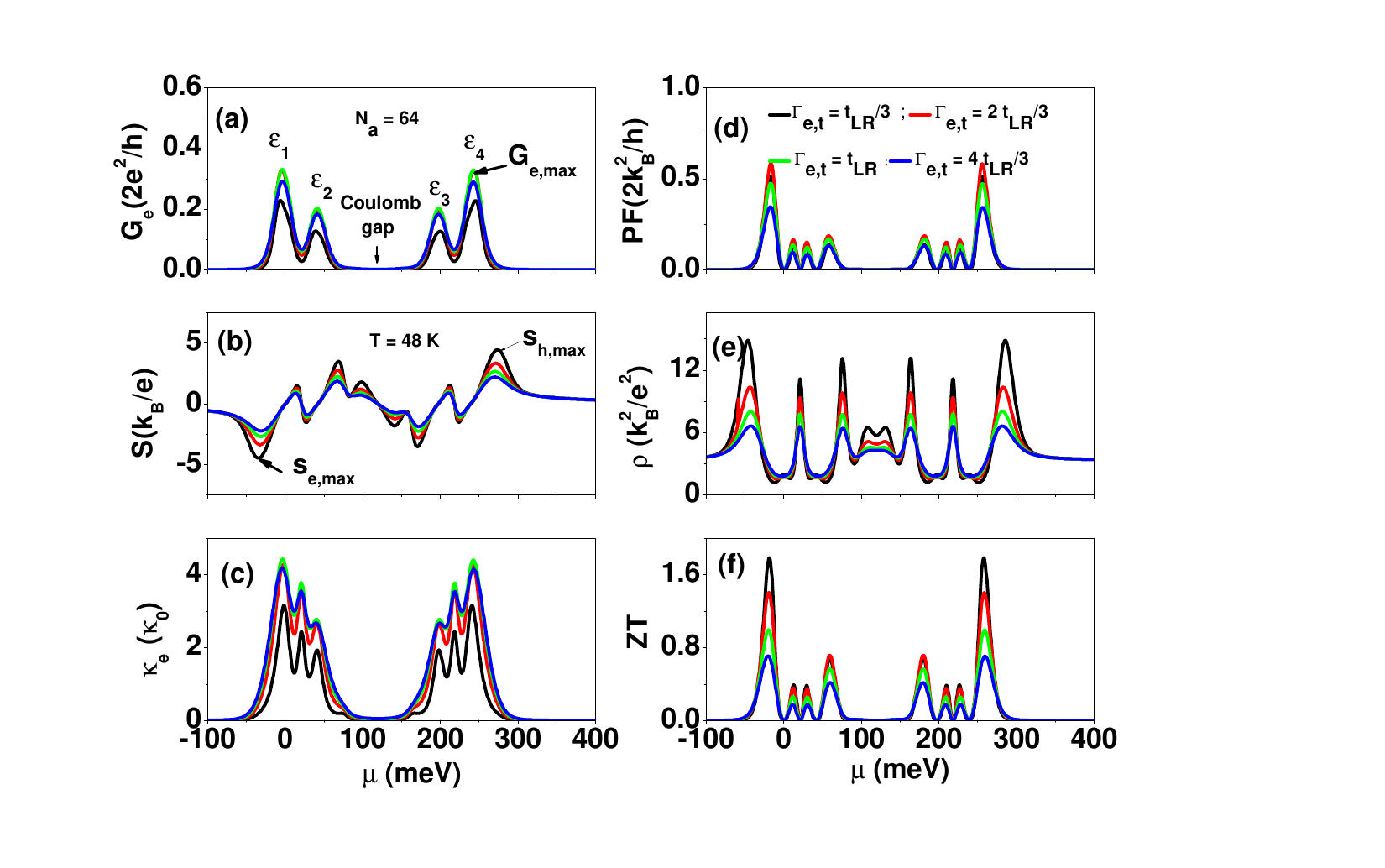}
\caption{(a) Electrical conductance $G_e$, (b) Seebeck coefficient
$S$ , (c) electron thermal conductance $\kappa_e$, (d) power
factor $PF$, (e) Lorenz number $\rho = \kappa_e/(TG_e)$ and (f)
figure of merit $ZT$ as functions of $\mu$ for various
$\Gamma_{e,t}$ values at $T = 48 K$. Other physical parameters are
the same as those of Fig. 6.}
\end{figure}

To effectively apply heat engines, it is crucial to clarify the
thermoelectric quantities over a wide temperature range. In Figure
8, we depict the behavior of $G_e$, $S$, $\kappa_e$, $PF$, $\rho$,
and $ZT$ as functions of $\mu$ for various temperatures at
$\Gamma_{e,t} = t_{LR}$. It can be observed that the magnitude of
$G_e$ reduces with increasing temperature. When the temperature is
at $200 K$ and $324 K$, the inter-site Coulomb interactions
responsible for the structure in $G_e$ spectra are washed out. In
Figure 8(b), the sophisticated spectra of $S$ become an N-shaped
curve at room temperature. In Figure 8(c), the highest $\kappa_e$
occurs at the middle Coulomb gap when the temperature is at $240
K$ and $324 K$. This behavior can be understood using the
expression for electron thermal conductance
$\kappa_e=\frac{1}{T}({\cal L}_2-{\cal L}^2_1/{\cal
L}_0)=\frac{{\cal L}_2}{T}-S^2G_eT$. When $\mu$ is located at the
middle Coulomb gap, electron-hole balance requires $S$ to be close
to zero. Therefore, $\kappa_e$ is dominated by $\frac{{\cal
L}_2}{T}$, which generally has a significant contribution at high
temperature based on the thermionic procedure where $\mu$ does not
align with resonant channels. This explains why $\kappa_e$ reaches
its maximum value in the middle Coulomb gap. The maximum power
factors are $PF=0.476$ and $PF=0.274$ for $T=48K$ and $T=324K$,
respectively. These values are lower than the maximum $PF$ values
in Figure 4(c), attributed to the reduction of $G_e$ resulting
from Coulomb blockade. As seen in Figure 8(e),
$\rho=\kappa_e/(TG_e)$ displays temperature-dependent behavior.
Clearly, the reduction of $G_e$ also suppresses the maximum $ZT$
values. Nevertheless, the maximum $ZT_{h(e),max}= 1.6$ at $T = 324
K$ still reaches the $80\%$ of $ZT_{h(e),max}= 1.966$ shown in
Fig. 4.

\begin{figure}[h]
\centering
\includegraphics[angle=0,scale=0.3]{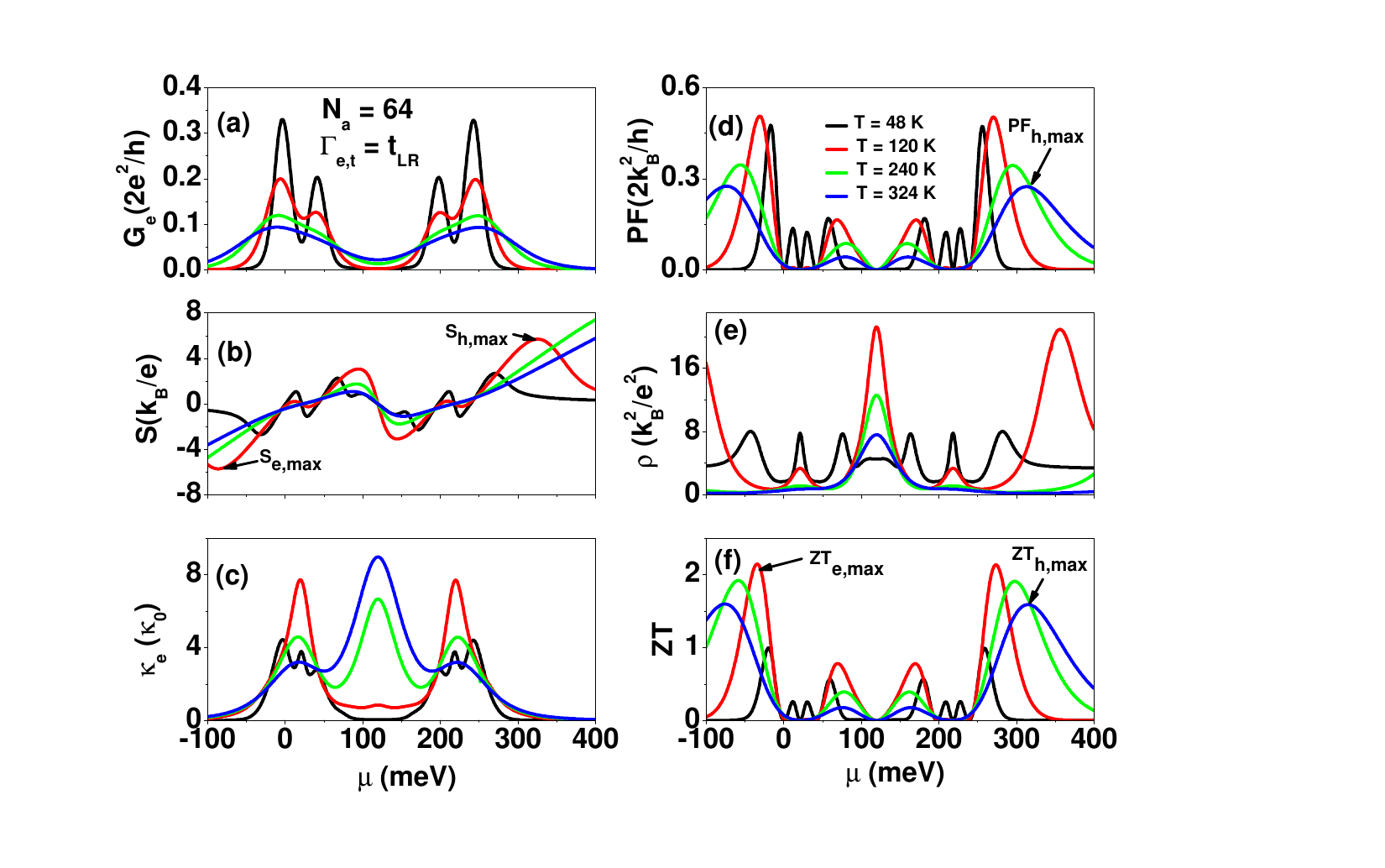}
\caption{(a) Electrical conductance, (b) Seebeck coefficient, (c)
electron thermal conductance, (d) power factor, (e) Lorenz number
merit $\rho$, and (f) figure of merit $ZT$ as functions of $\mu$
for different temperature values of AGNRs at
$\Gamma_{e,t}=t_{LR}$. The physical parameters considered are the
same as those in Figure 6. }
\end{figure}

\subsubsection{Nonlinear response regime}

Recent experimental studies have reported negative differential
conductance (NDC) in charge transport through the AGNR
heterostructure tunneling junction
[\onlinecite{Jacobse},\onlinecite{Mangnus}]. However, the
underlying mechanism for NDC at low bias remains unclear
[\onlinecite{Jacobse},\onlinecite{Mangnus}]. In this study, we
propose a novel mechanism for NDC resulting from the inter-zigzag
edge Coulomb interactions $U_1$ (or inter-site Coulomb
interactions). By utilizing Eqs. (5) and (6), we calculate the
tunneling current as a function of applied bias $V_{bias}$ for
different temperatures, as shown in Fig. 9(a) and 9(b). We have
considered $U_1$ values of 42 meV and 0 meV for Fig. 9(a) and Fig.
9(b), respectively. In the forward (reversed) bias, the tunneling
current is determined by the transmission coefficient of ${\cal
T}^{2-site}_{LR}(\varepsilon)$ (${\cal
T}^{2-site}_{RL}(\varepsilon)$). As seen in Fig. 9(a), at low
temperature (T = 12 K), the tunneling current in the low bias
region decreases with increasing $V_{bias}$. This intriguing
behavior demonstrates the NDC phenomenon, which is absent at
higher temperatures.

In Fig. 9(b), the NDC behavior disappears when we artificially set
the inter-site Coulomb interactions to zero ($U_1=0$) and only
take into account the intra-site Coulomb interactions ($U_0 = 155$
meV). The plateau of the tunneling current is attributed to the
intra-site Coulomb interaction of $U_0$, as $U_1$ is zero. It is
worth noting that in Fig. 9(a) and Fig. 9(b), we do not consider
the effect of bias-dependent orbital offset. Therefore, such NDC
characteristics are not a result of off-resonant channels
introduced by applied bias [\onlinecite{Mangnus}]. To clarify the
NDC behavior shown in Fig. 9(a), we have added two curves in Fig.
9(a) which were calculated considering the $P_1$ and $P_3$
channels of Eq. (6), respectively. The $P_3$ channel plays an
important role for $|V_{bias}|<50$ mV. However, it should be noted
that the tunneling current is dominated by the $P_1$ channel when
$|V_{bias}|>70$ mV. From the curves of $P_1$ and $P_3$, we
understand that the NDC behavior of the tunneling current is due
to the reduction of the probability of the $P_3$ channel. In Fig.
B.1, we provide the single-particle occupation number and two-site
two-particle correlation functions, such as the inter-site triplet
states $\langle n_{R,\sigma}n_{L,\sigma}\rangle$ ($\langle
n_{L,\sigma}n_{R,\sigma}\rangle$) and singlet states $\langle
n_{R,-\sigma}n_{L,\sigma}\rangle$ ($\langle
n_{L,-\sigma}n_{R,\sigma}\rangle$), which determine the
probabilities of the resonant channels $P_1$ and $P_3$ in ${\cal
T}^{2-site}_{LR}(\varepsilon)$ (${\cal
T}^{2-site}_{RL}(\varepsilon)$).

\begin{figure}[h]
\centering
\includegraphics[angle=0,scale=0.3]{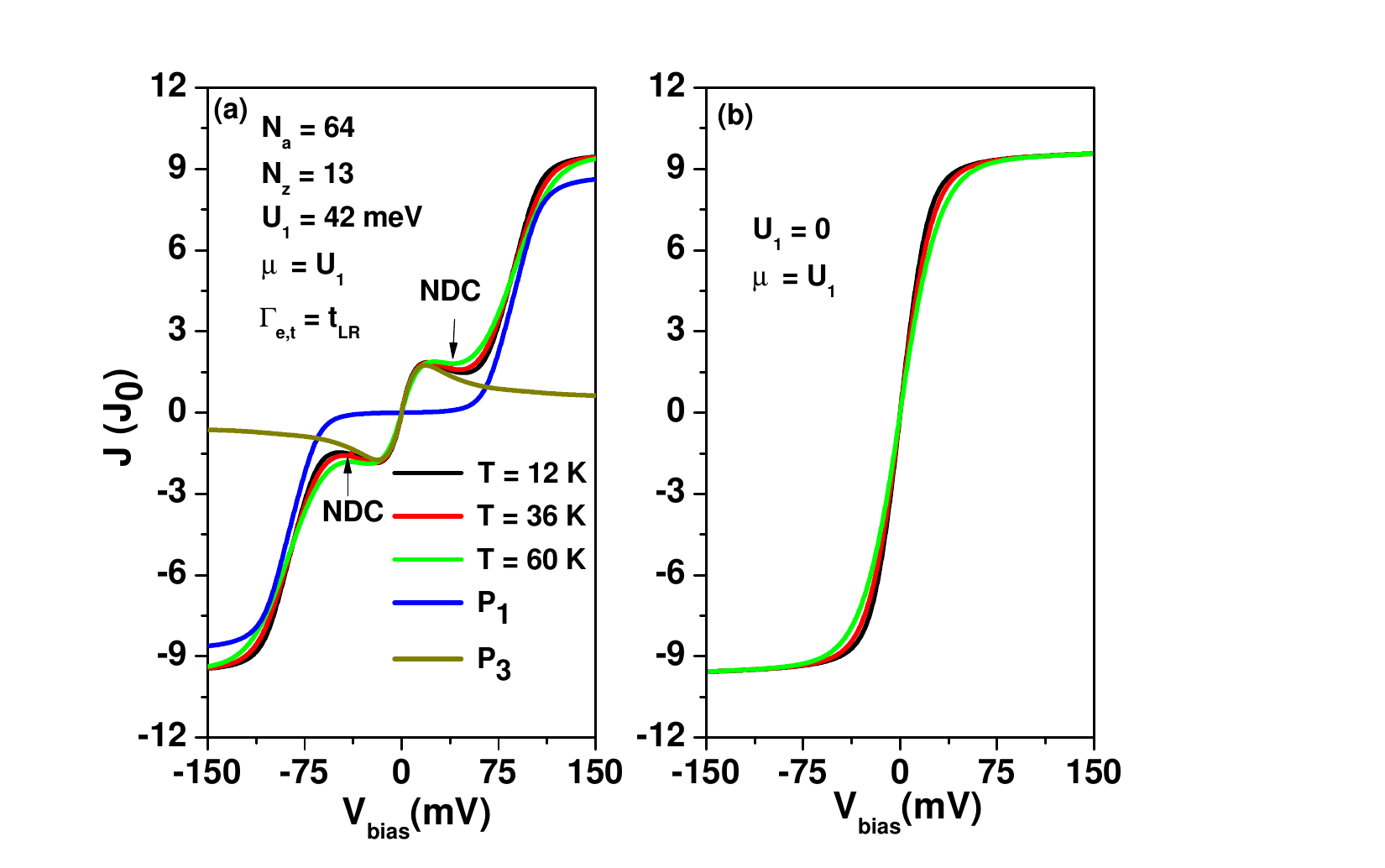}
\caption{ Tunneling current ($J$) as a function of applied bias
for various temperatures at $\Gamma_{e,t} = t_{LR}$, with $t_{LR}
= 7.13$ meV and $U_0 = 155$~meV. (a) shows the case where $U_1=
42$ meV, while (b) shows the case where $U_1 = 0$. We have also
set $E_L = E_R = 0$, and $\mu = U_1$. Additionally, the tunneling
current calculated by $P_1$ and $P_3$ channels at $T = 12$ K are
included in (a). The tunneling current is reported in units of
$J_0 = 0.773$~nA.}
\end{figure}

The electronic transport behavior can be significantly impacted by
the properties of the contact between AGNRs and the electrodes
[\onlinecite{DRizzo},\onlinecite{DJRizzo},\onlinecite{Joost}-\onlinecite{Su}].
Therefore, it is important to investigate the effect of the
parameter $\Gamma_{e,t}$ on the NDC behavior. In Fig. 10(a), we
present the tunneling current for different values of
$\Gamma_{e,t}$ at $T=12~K$, where the variation of $\Gamma_{e,t}$
corresponds to finite AGNRs or AGNR heterostructures with
different contact properties. To analyze the effect of the applied
bias on the orbital offset, we have adopted the method of Ref.
[\onlinecite{Pedersen}], where the bias-dependent energy level of
TSs is given by $E_{L(R)}=\eta_{L(R)}eV_{bias}$ in Fig. 10(b). As
shown in Fig. 10(a), the tunneling current increases as
$\Gamma_{e,t}$ increases, whereas the NDC region is reduced. Fig.
10(b) shows that the tunneling current is significantly suppressed
in the region of $|V_{bias}| > 50$ mV when the bias-dependent
orbital offset is taken into account. We have observed that a
second NDC region appears at high applied bias for $\eta = 0.16$.
This finding is consistent with the result presented in Figure
4(h) of the reference [\onlinecite{Mangnus}].

\begin{figure}[h]
\centering
\includegraphics[angle=0,scale=0.3]{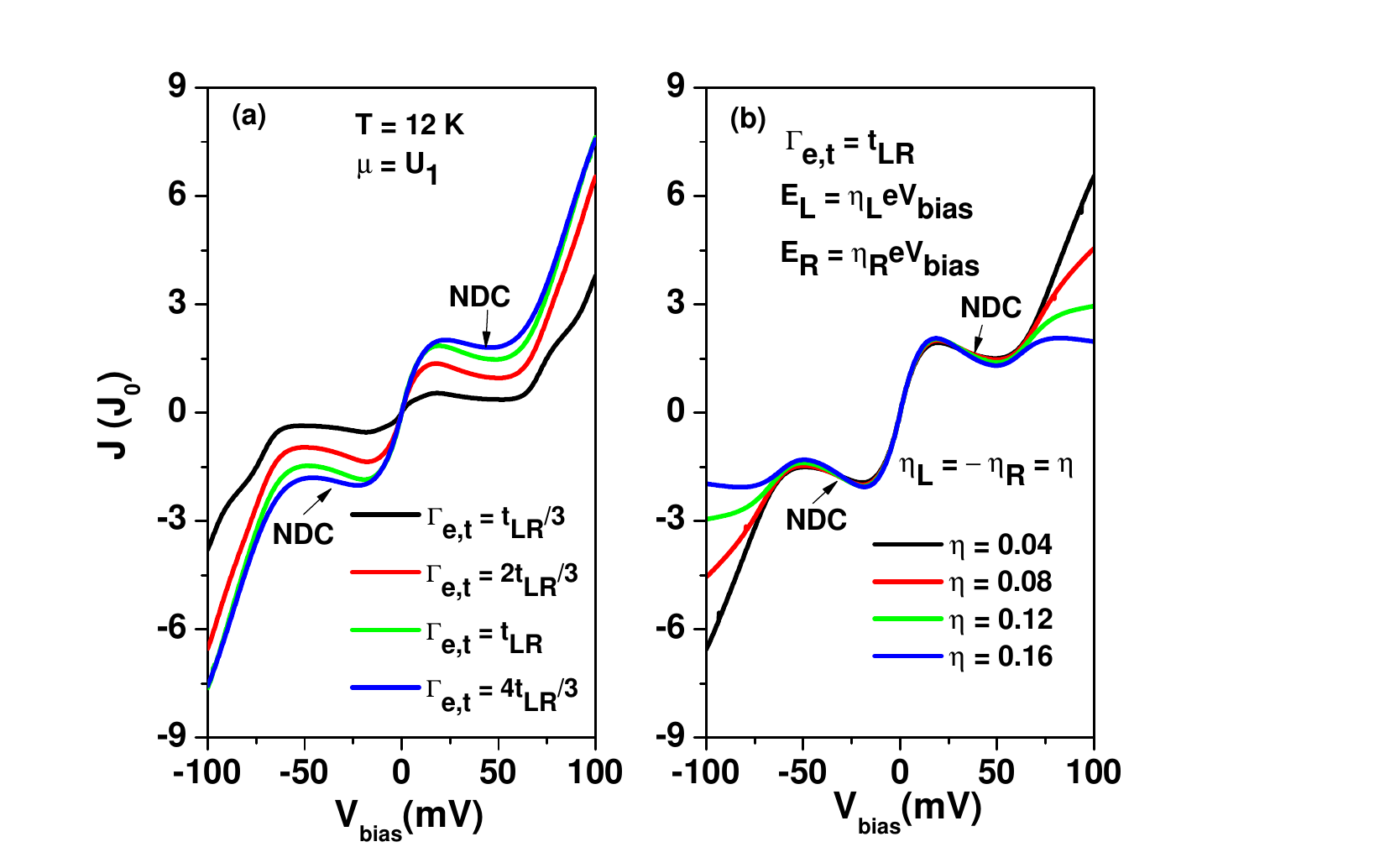}
\caption{(a) Tunneling current as a function of $V_{bias}$ for
various $\Gamma_{e,t}$ at $T = 12 K$ and $E_L = E_R = 0$, and (b)
tunneling current as a function of $V_{bias}$ for various $\eta $
values at $T = 12 K$, $\Gamma_{e,t}=t_{LR}$ and
$E_{L(R)}=\eta_{L(R)} eV_{bias}$. We have set $\mu = U_1 = 42$
meV. Other physical parameters are the same as those of Fig. 9.}
\end{figure}

Up until now, our calculations have assumed a symmetrical contact
junction with $\Gamma_{e,L}=\Gamma_{e,R}$. However, in Fig. 11, we
present tunneling current results for $\Gamma_{e,L} \neq
\Gamma_{e,R}$. In addition to NDC, we have also observed current
rectification behavior. The maximum current values in the low
forward and reversed applied bias are denoted as $J_{F,1}$ and
$J_{R,1}$, respectively. The ratio of $|J_{R,1}/J_{F,1}|$
increases with increasing $\Gamma_{e,R}$ when $\Gamma_{e,L}= 3$
meV. It reaches 1.91 for $\Gamma_{e,R}=12$~meV. As discussed in
Fig. 9, $J_{F,1}$ and $J_{R,1}$ are primarily attributed to the
$P_3$ channel of Eq. (6). The probability weight of $P_3$ is
primarily determined by the single-particle occupation number
$N_{R,\sigma} = \langle n_{R,\sigma}\rangle $
($N_{L,\sigma}=\langle n_{L,\sigma}\rangle$) and the two-site
singlet state $\langle n_{R,-\sigma} n_{L,\sigma}\rangle$
($\langle n_{L,-\sigma}n_{R,\sigma}\rangle$). When $|V_{bias}| <
150$ mV, the intra-site two-particle occupation number $\langle
n_{R,-\sigma}n_{R,\sigma}\rangle$ ($\langle
n_{L,-\sigma}n_{L,\sigma}\rangle$) and two-site three-particle
correlation functions are negligible. In the case of $\Gamma_{e,L}
=\Gamma_{e,R}$, we have $N_{L,\sigma}(V_{bias}) =
N_{R,\sigma}(-V_{bias})$. In contrast, if $\Gamma_{e,L} <
\Gamma_{e,R}$, then $N_{L,\sigma}(V_{bias}) \neq
N_{R,\sigma}(-V_{bias})$. Meanwhile, the two-site singlet state
(2-site-S) in the reversed bias region is smaller than that in the
forward bias region. Therefore, we observe that $|J_{R,1}|$ is
larger than $J_{F,1}$. Note that if the inter-site Coulomb
interaction $U_1$ is set to zero, the current rectification is
significantly reduced as $\Gamma_{e,L} \neq \Gamma_{e,R}$. This
suggests that the inter-site Coulomb interactions play a crucial
role in determining the behavior of negative differential
conductance (NDC) and current rectification.

\begin{figure}[h]
\centering
\includegraphics[angle=0,scale=0.3]{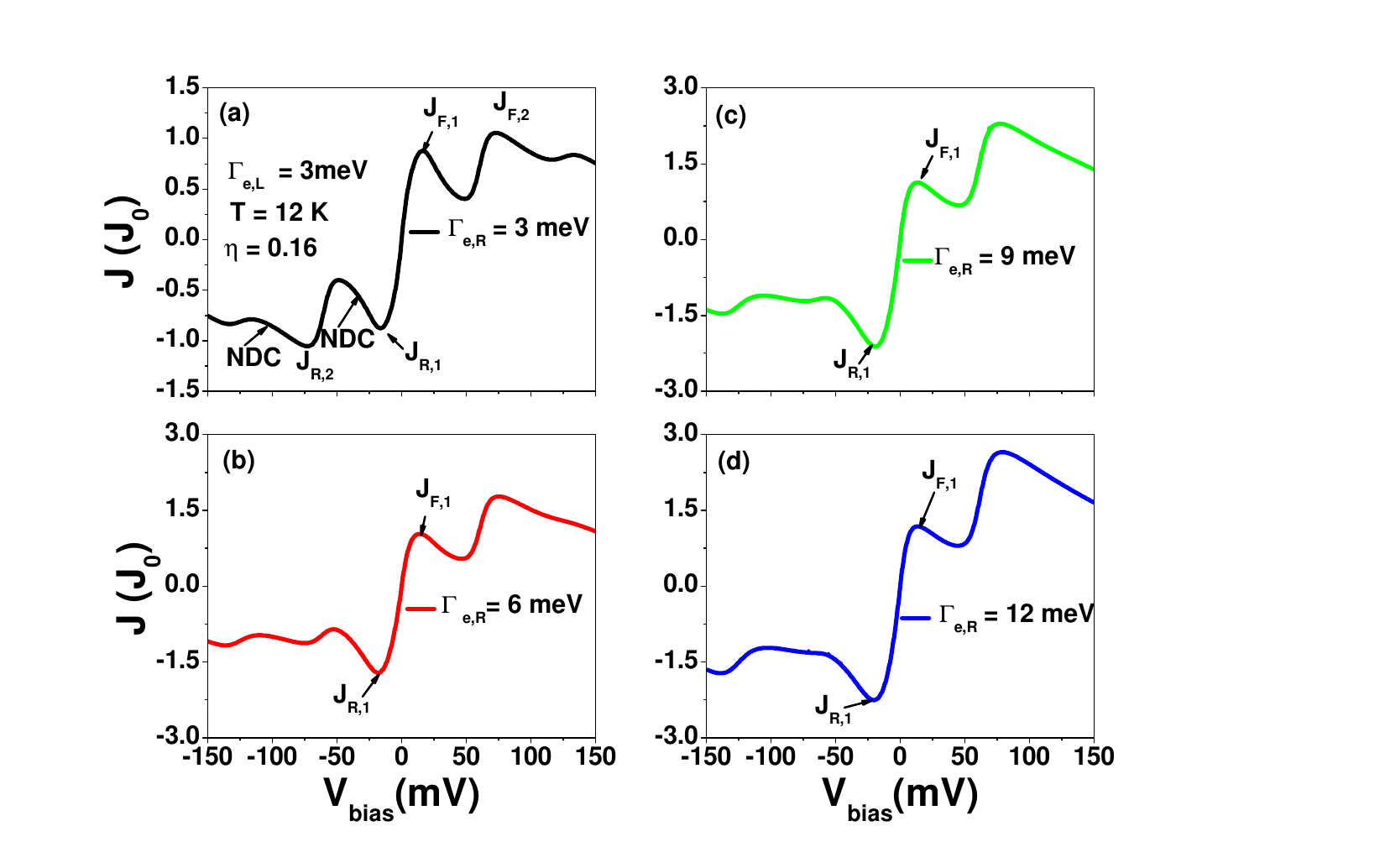}
\caption{Tunneling current as a function of $V_{bias}$ for various
$\Gamma_{e,R}$ values at $\Gamma_{e,L}= 3$ meV, $T = 12 K$ and
$\eta_L=-\eta_R=0.16$. $\mu = U_1 = 42$ meV. (a) $\Gamma_{e,R} =
3$~meV, (b) $\Gamma_{e,R} = 6$~meV, (c) $\Gamma_{e,R} = 9$~meV and
(\textbf{d}) $\Gamma_{e,R} = 12$~meV. Other physical parameters
are the same as those of Fig. 9.}
\end{figure}

\subsection{Tunneling Current through TSs of 9-7-9 AGNR Heterostructures under Pauli Spin Blockade Configuration}
Because the topological states (TSs) of 9-7-9 AGNR heterostuctures
locates at the interfaces between 9-7 junctions, which can be very
far way from the contacted electrodes based on the design (see
Appendix A.), it is possible to lay-out the two gate electrodes to
modulate the energy levels of TSs and set them in the Pauli spin
blockade (PSB) configuration. We present the calculated tunneling
current through the TSs of 9-7-9 AGNR heterostructures in the PSB
configuration as a function of $V_{bias}$ for various
$\Gamma_{e,L} = \Gamma_{e,R} = \Gamma_{e,t}$ values at $T = 12 K$
in Fig. 12, where each AGNR segment has 8 unit cells. Such TSs
have $U_0 = 125$~meV, $U_1 = 49$~meV and $t_{LR} = 7.54$~meV. As
seen in Fig. 12, a remarkable current rectification behavior is
observed in the PSB configuration at symmetrical tunneling rate
$\Gamma_{e,L}=\Gamma_{e,R}$. The ratio of $|J_{R,max}/J_{F,max}|$
are 4.242, 2.908, 2.477 and 2.286 for $\Gamma_{e,t} = t_{LR}/3$,
$\Gamma_{e,t} = 2 t_{LR}/3$, $\Gamma_{e,t} = t_{LR}$ and
$\Gamma_{e,t} = 4 t_{LR}/3$, respectively.

\begin{figure}[h]
\centering
\includegraphics[angle=0,scale=0.3]{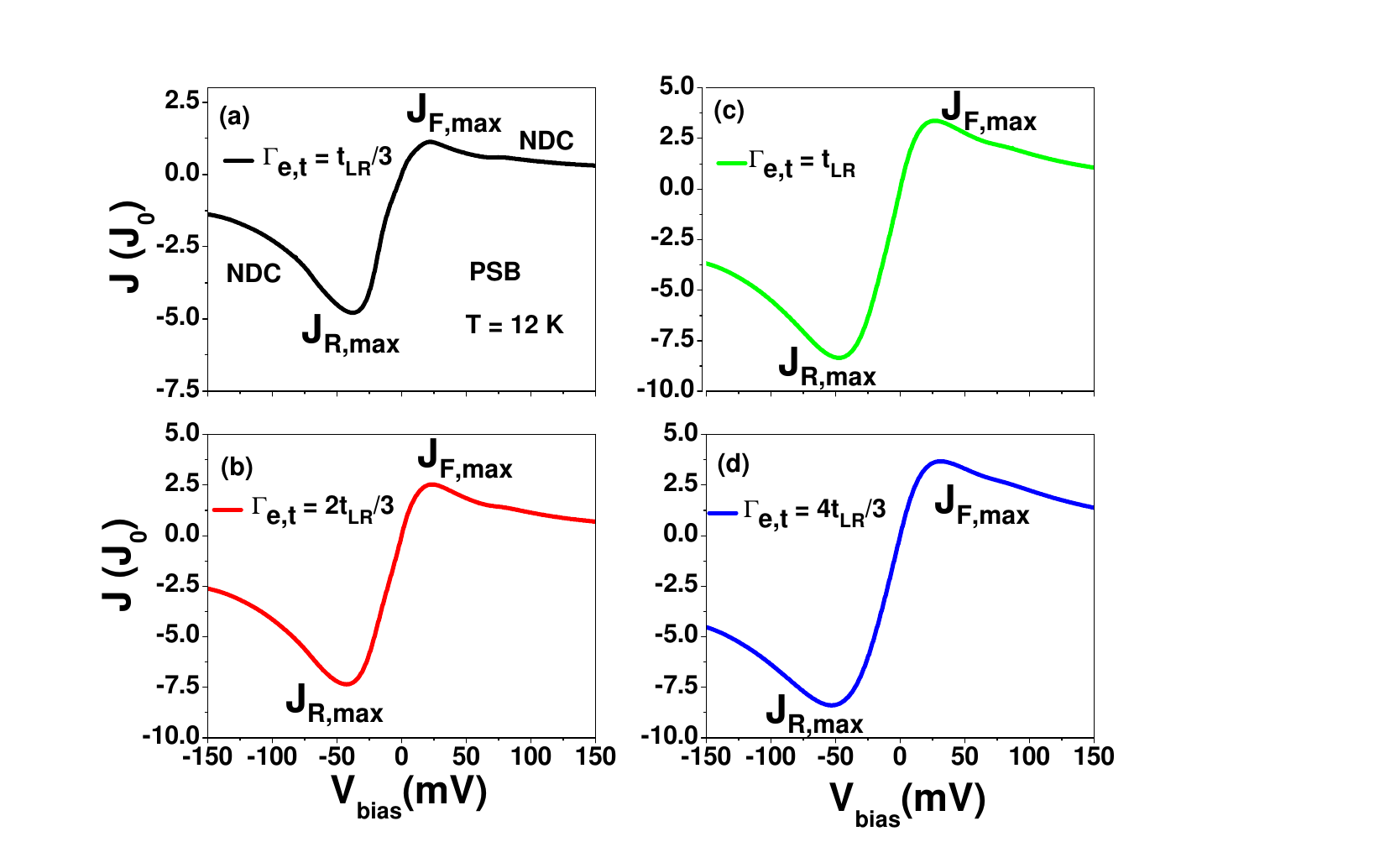}
\caption{Tunneling current through the topological states (TSs) of
9-7-9 AGNR heterostructures in the Pauli spin blockade
configuration as a function of $V_{bias}$ for various
$\Gamma_{e,t}$ values at $T = 12 K$. The energy levels of the left
and right TSs are given by $E_L = -U_1+\eta_L eV_{bias}$ and $E_R
= -U_0+\eta_R eV_{bias}$, respectively, where $\mu = 0 $ meV and
$\eta_L = -\eta_R = 0.16$. (a) $\Gamma_{e,t} = t_{LR}/3$, (b)
$\Gamma_{e,t} = 2 t_{LR}/3$, (c) $\Gamma_{e,t} = t_{LR}$ and (d)
$\Gamma_{e,t} = 4 t_{LR}/3 $. The parameters used are $U_0 = 125$
meV, $U_1 = 49$ meV, and $t_{LR} = 7.54$ meV, which correspond to
AGNR hterostructures formed by 8 u.c. segments.}
\end{figure}

To understand the rectification behavior shown in Figure 12, we
first note that under PSB, the channel $P_2$ in Eq. (5) behaves as
a resonant channel in the forward applied bias, and the tunneling
current is mainly contributed by $P_2$. We therefore focus on the
transmission coefficient of $P_2$ in the PSB configuration, which
can be expressed as:

\begin{small}
\begin{eqnarray}
& &{\cal T}_{PSB}(\varepsilon) \\ &=&\frac{P_{PSB} }
{|(\varepsilon-E_L-U_1+i\Gamma_{e,L})(\varepsilon-
E_R-U_0+i\Gamma_{e,R})-t^2_{LR}|^2}, \nonumber
\end{eqnarray}
\end{small}

Here, the probability of resonant channel is given by
$P_{PSB}=N_{R,\sigma}-\langle n_{R,\sigma} n_{L,\sigma}\rangle
-\langle n_{R,-\sigma} n_{R,\sigma}\rangle+\langle n_{R,-\sigma}
n_{R,\sigma} n_{L,\sigma}\rangle$ ($P_{PSB}=N_{R,\sigma}- \langle
n_{L,\sigma}n_{R,\sigma}\rangle -\langle
n_{L,-\sigma}n_{R,\sigma}\rangle+\langle
n_{L,-\sigma}n_{L,\sigma}n_{R,\sigma}\rangle $) in the forward
(reversed) applied bias. In the small applied bias region
$|V_{bias}|\le 50$~mV, the resonant channels are almost unaffected
by bias-dependent orbital offset, and $P_{PSB}$ is the main factor
determining the magnitude of the tunneling current. For
$|V_{bias}|\le 150$~mV, only single occupation number and
inter-site singlet and triplet states are important. We show these
correlation functions in Figure 13 to clarify the bias-dependent
$P_{PSB}$.

As seen in Figure 13(a), charge transport in the PSB region is a
two-electron process because of the total electron number $1.5\le
N_T \le 2$. In the forward applied bias, $N_{R,\sigma}$ and
$N_{L,\sigma}$ are in discharging and charging processes,
respectively, and their values reach 0.5 at $V_{bias}=150$~mV. Due
to the large probability of inter-site triplet state (see the
curve of 2-site-T), the $P_{PSB}$ is degraded. In contrast,
$N_{R,\sigma}$ and $N_{L,\sigma}$ in the reversed applied bias are
in the charging and discharging processes, respectively. The
highly enhanced $N_{R,\sigma}$ and suppressed inter-site
correlation functions (see 2-site-T and 2-site-S) result in a
large $P_{PSB}$. This explains the significant current
rectification observed in Figure 13(a). With increasing
$\Gamma_{e,t}$, the magnitude of 2-site-T is degraded for
$V_{bias} > 0$, leading to an enhancement of $J_{F,max}$ and a
reduction of $|J_{R,max}/J_{F,max}|$ in Figure 13. In addition, we
observe an interesting behavior of phase transformation. The merge
together 2-site-T and 2-site-S curves in the reversed applied bias
region is splitting into two curves in the forward applied bias.

\begin{figure}[h]
\centering
\includegraphics[angle=0,scale=0.3]{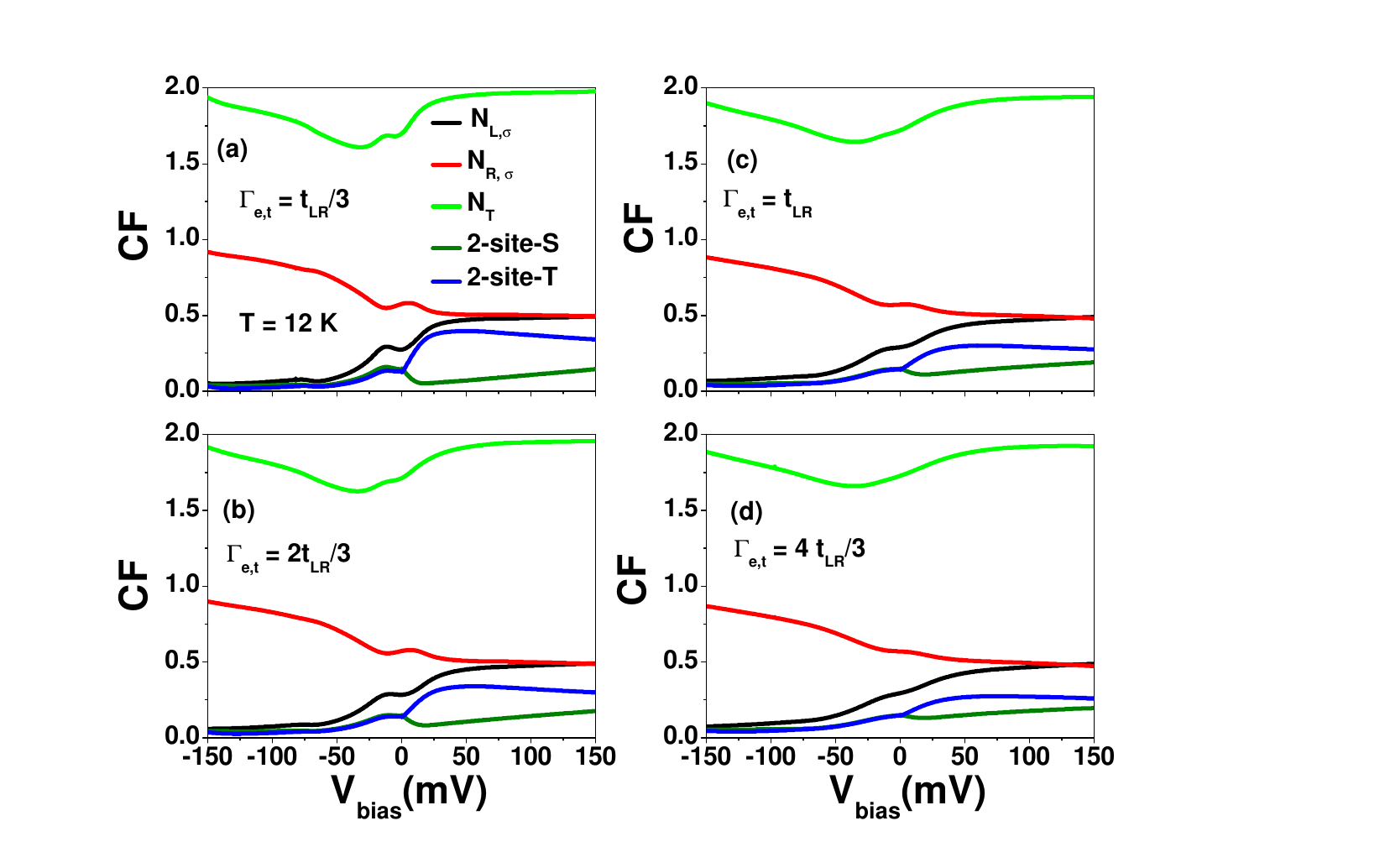}
\caption{Correlation functions ($CF$) of 9-7-9 AGNR
heterostructures in the PSB for various $\Gamma_{e,t}$ values at
$T = 12 K$. (a) $\Gamma_{e,t} = t_{LR}/3$, (b) $\Gamma_{e,t} = 2
t_{LR}/3$, (c) $\Gamma_{e,t} = t_{LR}$ and (d) $\Gamma_{e,t} = 4
t_{LR}/3 $. Other physical parameters are the same as those of
Fig. 12.}
\end{figure}

Finally, we examine SCTSs in the PSB configuration for weak
coupling strength of $t_{LR}$ values. In Fig. 14, we present (a)
single particle occupation numbers $N_{j,\sigma}$, (b) inter-site
two particle correlation functions, and (c) tunneling current of
9-7-9 AGNR heterostructures as functions of $V_{bias}$ at $T = 12
K$ , $t_{LR} = 0.88$~meV, $\Gamma_{e,L} = t_{LR}/3$ and
$\Gamma_{e,R} = 3t_{LR}$. To achieve the small coupling parameter
of $t_{LR} = 0.88$~meV, we use 12 u.c. segments to form 9-7-9 AGNR
heterostructures. We can ignore the bias-dependent orbital offset
between TSs since the wave functions of TSs are far away from the
electrodes [\onlinecite{Pedersen}]. As seen in Fig. 14 (b), the
maximum value of the 2-site-T curve ($0.486$) and the minimum
value of the 2-site-S curve (nearly zero) occur in the forward
applied bias. Compared to the results of Fig. 13, SCTSs are highly
occupied by the inter-site triplet states in the forward applied
bias for a weak $t_{LR}$. Since we ignore the bias-dependent
orbital offset between TSs, the vanishingly small current of $J_F$
is generated by SCTS highly occupied by the triplet states in Fig.
14 (c). The significant current rectification in the PSB
configuration shown in Figure 14 is valuable in spin-current
conversion devices[\onlinecite{Ono}].

\begin{figure}[h]
\centering
\includegraphics[angle=0,scale=0.3]{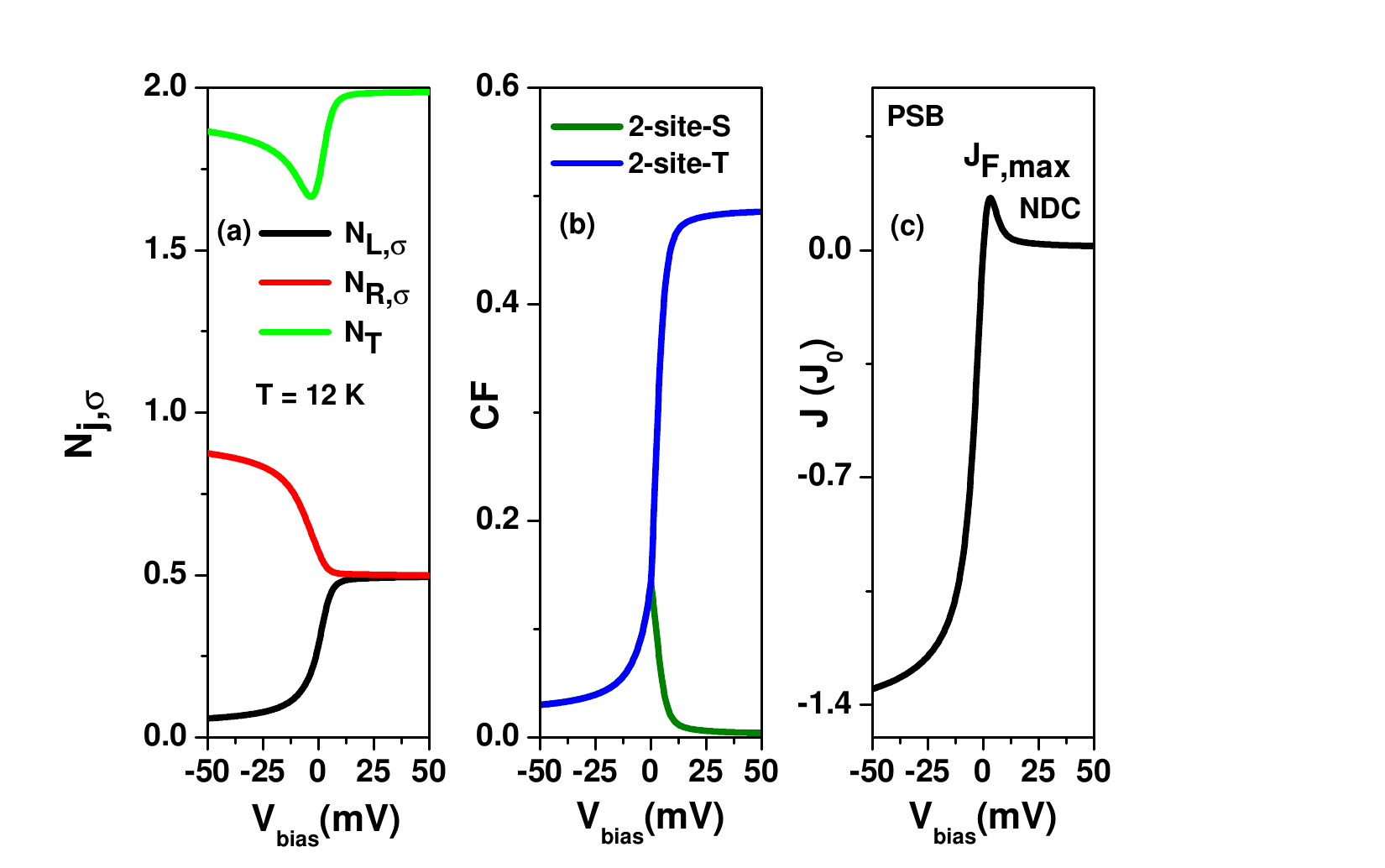}
\caption{ (a) Single particle occupation numbers $N_{j,\sigma}$,
(b) inter-site two particle correlation functions (CF), and (c)
tunneling current of 9-7-9 AGNR heterostructures as functions of
$V_{bias}$ for a small $t_{LR}=0.88$~meV at $T = 12 K$,
$\Gamma_{e,L}= t_{LR}/3$, and $\Gamma_{e,R}= 3t_{LR}$. We realize
the small $t_{LR} = 0.88$~meV using 12 u.c. segments to form 9-7-9
AGNR heterostructures. Their intra-site and inter-site Coulomb
interactions are $U_0 = 111$~meV and $U_1 = 36.97$~meV. The energy
levels of TSs are $E_L = - U_1$ and $E_R = - U_0$.}
\end{figure}

\section{Conclusion}
In summary, this study provides an in-depth investigation of the
charge transport properties of two distinct graphene nanoribbon
(GNR) structures: the end zigzag edges of AGNRs and the
topological states of 9-7-9 AGNR heterostructures. Our findings
demonstrate that 9-7-9 AGNR heterostructures with deep topological
states (TSs) exhibit superior characteristics, such as wave
functions of TSs far away from the electrodes, making them more
suitable for low-power quantum devices. The~two-site model with
effective tunneling rates provides an excellent description of the
electrical conductance spectra of the serially coupled topological
states (SCTSs). Additionally, we analyzed the Coulomb blockade
effect on the thermoelectric coefficients and tunneling current of
the zigzag edge states using the two-site Hubbard model. Our
results show that electron Coulomb interactions have a more
significant impact on electrical conductance than on the Seebeck
coefficient. The~Lorenz number of the zigzag edge states does not
satisfy the Wiedemann--Franz law due to their localized
characteristic.

We observed negative differential conductance (NDC) in the
nonlinear response regime of the tunneling current, attributed to
inter-zigzag edge Coulomb interactions. Additionally, we observed
current rectification behavior in the end zigzag edges of AGNRs
when asymmetrical contacted electrode junctions were present. The~
tunneling current through SCTSs formed by 9-7-9 AGNR
heterostructures in the Pauli spin blockade configuration exhibits
remarkable current rectification behavior, even in AGNR
heterostructures with symmetrical contacted electrodes. For~a weak
coupling parameter $t_{LR}$, the~forward tunneling current is
almost blocked due to the high occupation of SCTSs by two-electron
triplet states. This property is highly useful in spin-current
conversion devices. Overall, our study provides valuable insights
into the charge transport properties of TSs in finite AGNRs and
heterostructures, emphasizing the importance of considering
electron-electron interactions in understanding their~behavior.


{}

{\bf Acknowledgments}\\
{This work was supported by the Ministry of Science and Technology
(MOST), Taiwan under Contract No. MOST 107-2112-M-008-023MY2.}

\mbox{}\\
E-mail address: mtkuo@ee.ncu.edu.tw\\

\appendix
\numberwithin{figure}{section}
\section{9-7-9 AGNR heterostructures}
\numberwithin{figure}{section}

\numberwithin{equation}{section}

The results presented in Fig. 2(b) indicate that the wave function
of the left zigzag edge state has only a small overlap with that
of the right zigzag edge state for the case of $N_z=9$.
Consequently, it is difficult for electrons from the electrodes to
be transported through these localized states. Recently,
interesting topological states (TSs) have been identified in the
electronic structures of 9-7 AGNR heterostructures. In these
structures, the wave function of TS accumulates at the interface
between the 9-7 junction
[\onlinecite{Rizzo}--\onlinecite{DJRizzo}]. However, the charge
transport through these serially coupled TSs (SCTSs) has not been
clarified
[\onlinecite{Mangnus},\onlinecite{Joost},\onlinecite{Lopez}]. In
the appendix A, we not only further clarify their electronic
structures but also investigate how the contacted electrodes
influence the charge transport through the SCTSs.

\begin{figure}[h]
\centering
\includegraphics[angle=0,scale=0.3]{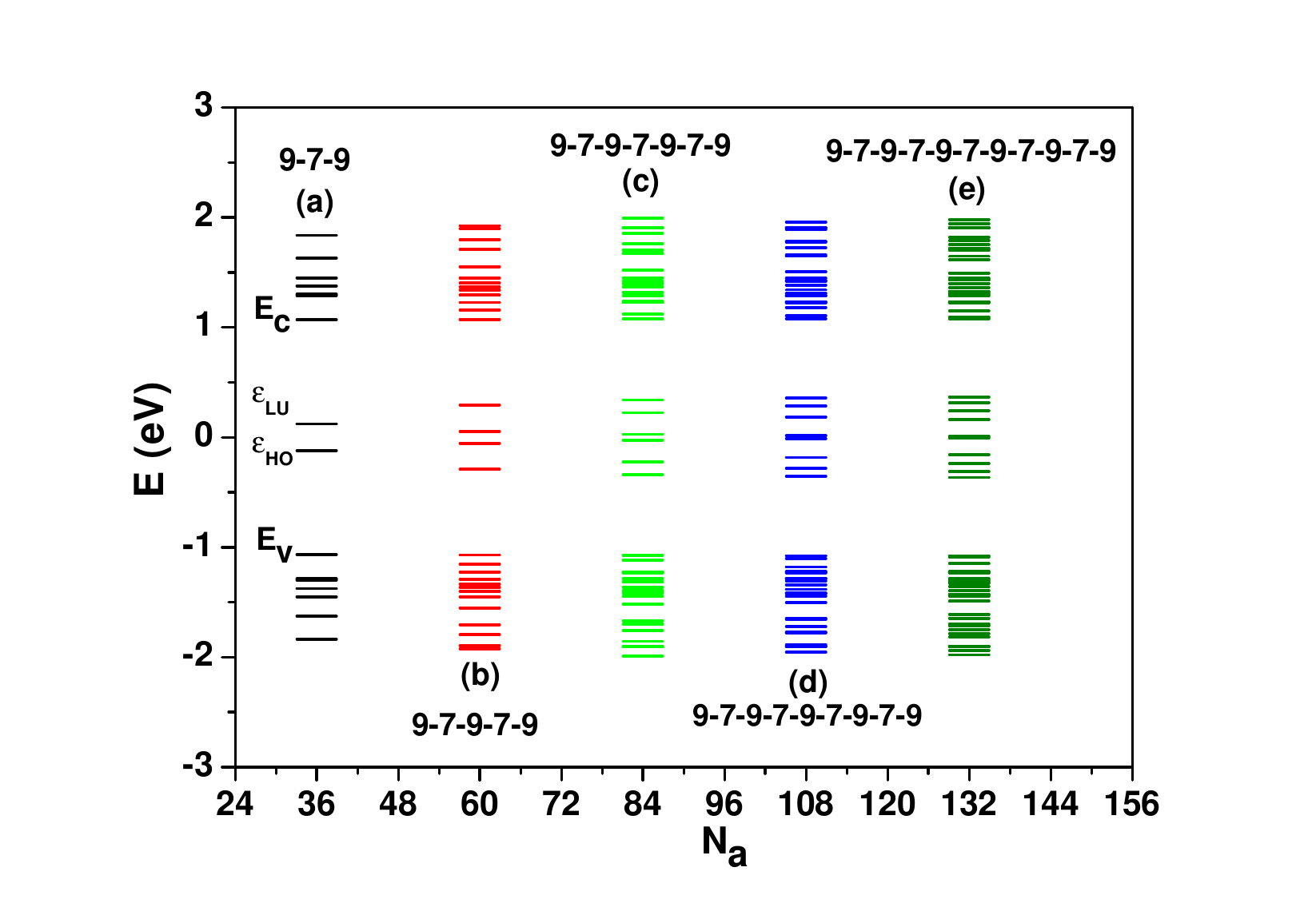}
\caption{Eigenvalues of AGNR heterostructures with varying numbers
of 7-AGNR segments. 9-7-9 AGNR heterostructures are formed by 3
u.c segments. (a) to (e) correspond to one to five 7-AGNR
segments, respectively, with an energy range of $|E| \leq 2$~eV.}
\end{figure}

As depicted in Fig. 1(b), the 9-7-9 AGNRs are formed by 3 u.c.
segments. The calculated eigenvalues of the AGNR heterostructures
for various periods are shown in Fig. A.1. As seen in the case of
(a), two peaks labeled by $\varepsilon_{HO}$ and
$\varepsilon_{LU}$ with a splitting energy of $|2t_{LR}| =
0.24312$~eV between $E_c$ and $E_v$ are observed. Here, $E_c$ and
$E_v$ denote the conduction subband minimum and valence subband
maximum, respectively. The charge density distribution of
$\varepsilon_{LU} = 0.12156$~eV is presented in Fig. 1(b), which
shows that the charge densities are very dilute for lattices close
to the end zigzag edges. Such behavior indicates that
$\varepsilon_{HO}$ and $\varepsilon_{LU}$ have weak coupling
strengths with the electrodes. As each 7-AGNR segment provides two
energy levels, there are 4, 6, 8, and 10 energy levels within the
band gap for (b), (c), (d), and (e), respectively, as shown in the
results. When the number of 7-AGNR segments approaches infinity,
the minibands are formed. To reveal such an interesting electronic
structures formed by TSs, we show the calculated electronic
structures of 9-7 AGNR superlattices (SLs) in Fig. A.2, where (a)
and (b) consider different 9-AGNR segment lengths at fixed 3 u.c.
7-AGNR segments. Two minibands are formed near charge neutrality
point (CNP), the tiny gaps in (a) and (b) are $76$~meV and
$68$~meV, respectively. We find that the electronic dispersions of
minibands can be well described by Su-Schrieffer-Heeger (SSH)
model with a closed expression of
$E_{SSH}(k)=\pm~\sqrt{t^2_1+t^2_2-2t_1t_2
cos(k~\pi/L)}$[\onlinecite{Groning},\onlinecite{DRizzo},
\onlinecite{Su}]. Using  $E_{SSH}(k)$, we can determine $t_1 =
0.129$~eV and $t_2 = 0.167$~eV in (a) and $t_1 = 0.118$~eV and
$t_2=0.084$~eV in (b). As seen in Fig. A. 2, the miniband widths
become narrow as the length of 9-AGNR segment increases. According
to the SSH model, the metal, semiconductor and insulator phases
are determined by the relationships between $t_1$ and $t_2$. Once
$t_1 = t_2$, the band gap is vanishingly small.

\begin{figure}[h]
\centering
\includegraphics[angle=0,scale=0.3]{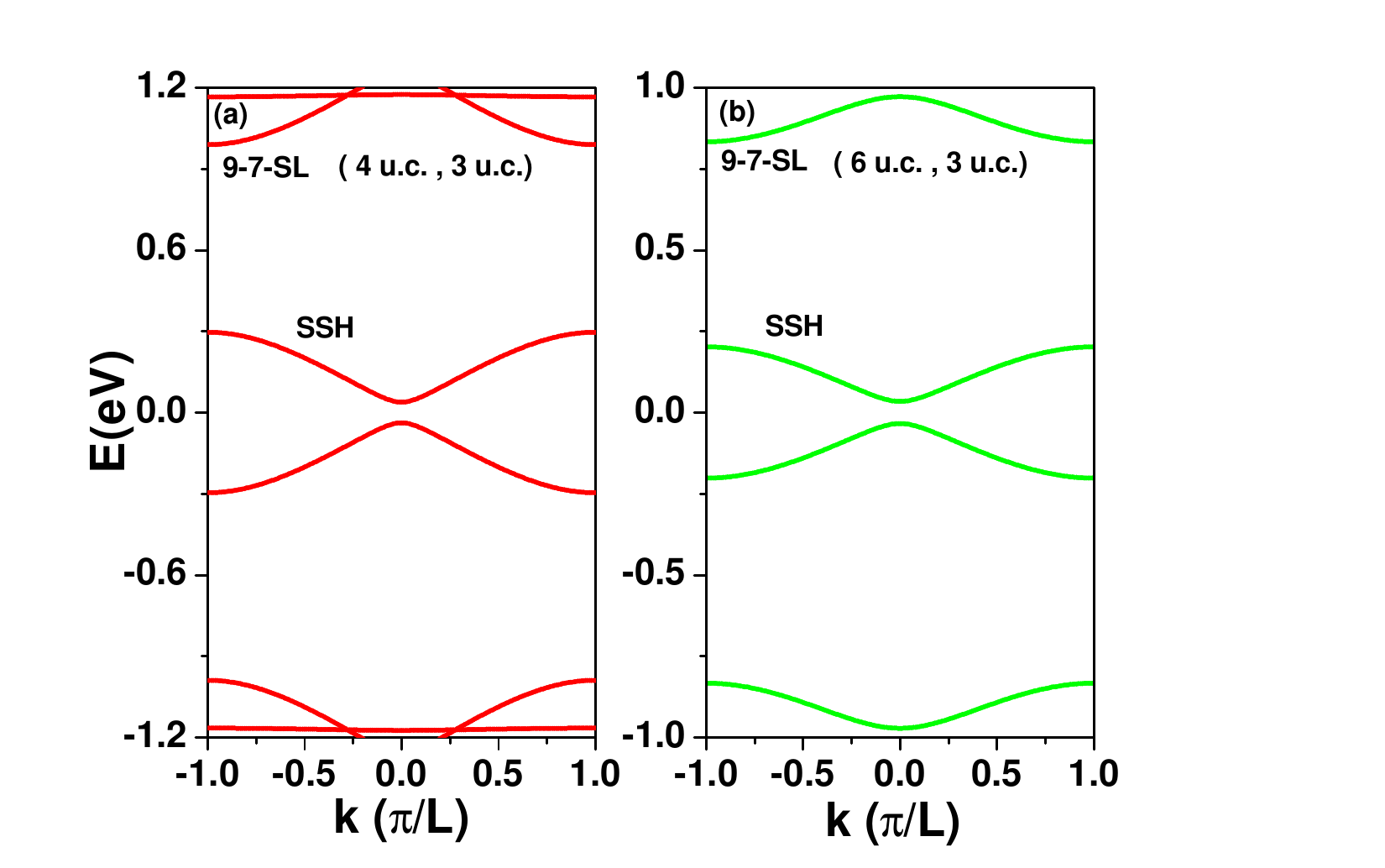}
\caption{ Electronic structures of 9-7 AGNR superlattices (SLs).
(a) and (b) correspond to 9-AGNR segments with 4 u.c. and 6 u.c.,
respectively, with an energy range of $|E| \leq 1.2$~eV. Note that
7-AGNR segment with 3 u.c. is fixed. $L$ is the length of super
unit cell. We have $L = 7$ u.c. and $L = 9$ u. c. in (a) and (b),
respectively.}
\end{figure}

The charge densities of the topological states (TSs) in 9-7-9 AGNR
heterostructures are distinct from those of the end zigzag edge
states found in finite AGNRs. The wave functions of the TSs are
located far away from the line-contacted electrodes, which allows
for the observation of charge transport through the TSs of 9-7-9
heterostructures even at large coupling strengths between the
zigzag edge carbon atoms and the electrodes (i.e., large
$\Gamma_t$ values). We calculated the electrical conductance of
6-unit cell (u.c.) 9-7-9 AGNRs for various $\Gamma_t$ values, and
the results are presented in Figure A.3. The splitting energy
between the highest occupied ($\varepsilon_{HO}$) and lowest
unoccupied ($\varepsilon_{LU}$) states is $|2t_{LR}|=44.4$ meV,
and their magnitudes and widths increase with increasing
$\Gamma_t$. Notably, even when $\Gamma_t=2.7$ eV, the two peaks of
$\varepsilon_{HO}$ and $\varepsilon_{LU}$ are still clearly
resolved. The excellent agreement between the results of Fig. A.3
and Eq. (6) (not shown here) is due to the TSs' localized wave
functions. Therefore, a 2-site Hubbard model is also suitable for
describing charge transport through the TSs of 9-7-9 AGNR
heterostructures.

\begin{figure}[h]
\centering
\includegraphics[angle=0,scale=0.3]{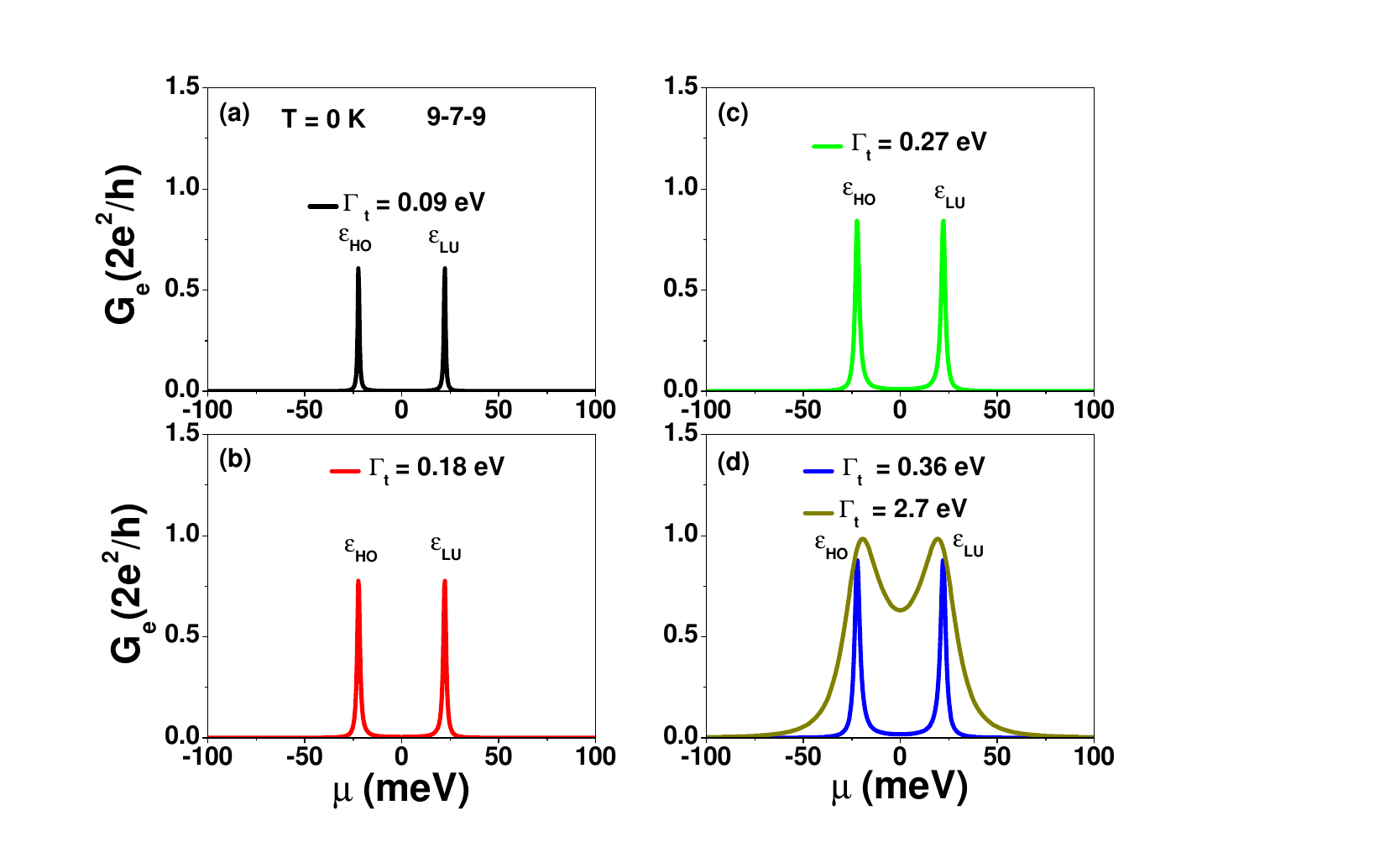}
\caption{Electrical conductance of 9-7-9 AGNR heterostructures
formed by 6 u.c. segments as functions of chemical potential
($\mu$) for various $\Gamma_t$ values at $T = 0 K$.}
\end{figure}

\section{Correlation functions of the end zigzag edges of AGNRs}
\numberwithin{figure}{section} \numberwithin{figure}{section}

In Figure 9, we illustrated the tunneling current of the end
zigzag edge states as a function of $V_{bias}$ for various
temperatures. The behavior of the tunneling current is determined
by the resonant channels and their probabilities. Since $E_L$ and
$E_R$ do not have a bias-dependent offset, the NDC behavior arises
from the bias-dependent probabilities. Therefore, it is essential
to understand the behavior of $P_n$ in Eq.(6). Figure B.1 displays
the single-particle occupation number $N_{L(R),\sigma}=\langle
n_{L(R),\sigma}\rangle$ and inter-site two-particle correlation
functions (CF) as functions of the applied bias ($V_{bias}$), with
and without inter-site Coulomb interactions. Due to the
symmetrical structure of the junction system, we have
$N_{L,\sigma}(V_{bias}) = N_{R,\sigma}(-V_{bias})$. The 2-site
singlet (2-site-S) state $\langle n_{R,-\sigma}
n_{L,\sigma}\rangle$ and 2-site triplet (2-site-T) state $\langle
n_{R,\sigma} n_{L,\sigma}\rangle$ (or $\langle n_{L,-\sigma}
n_{R,\sigma}\rangle$ and $\langle
n_{L,\sigma}n_{R,\sigma}\rangle$) also exhibit this symmetric
character with respect to $V_{bias}$. We analyzed their behavior
in the case of a forward bias situation ($V_{bias}>0$).
$N_{L(R),\sigma}$ has a finite value at zero bias ($V_{bias}=0$
mV) since $E_L$ and $E_R$ are below $\mu$. As $V_{bias}$
increases, $N_{L,\sigma}$ increases, but $N_{R,\sigma}$ decreases.
Eventually, $N_{R,\sigma}$ will saturate. The charge filling of
$E_R$ by the left electrode is attributed to the resonant channel
between $E_L$ and $E_R$. Since $N_{R,\sigma}$ is indirectly
charged by the left electrode, in the forward applied bias range,
$N_{R,\sigma}$ is smaller than $N_{L,\sigma}$. We found that the
probability of 2-site-S is larger than that of 2-site-T at finite
bias. The bias-dependent probabilities of $P_1$ and $P_3$
channels, which provide resonant energy levels, are plotted in
Fig. B.1(a). The decline behavior of the probability of $P_3$
explains the behavior of the tunneling current with NDC shown in
Fig. 9.

\begin{figure}[h]
\centering
\includegraphics[angle=0,scale=0.3]{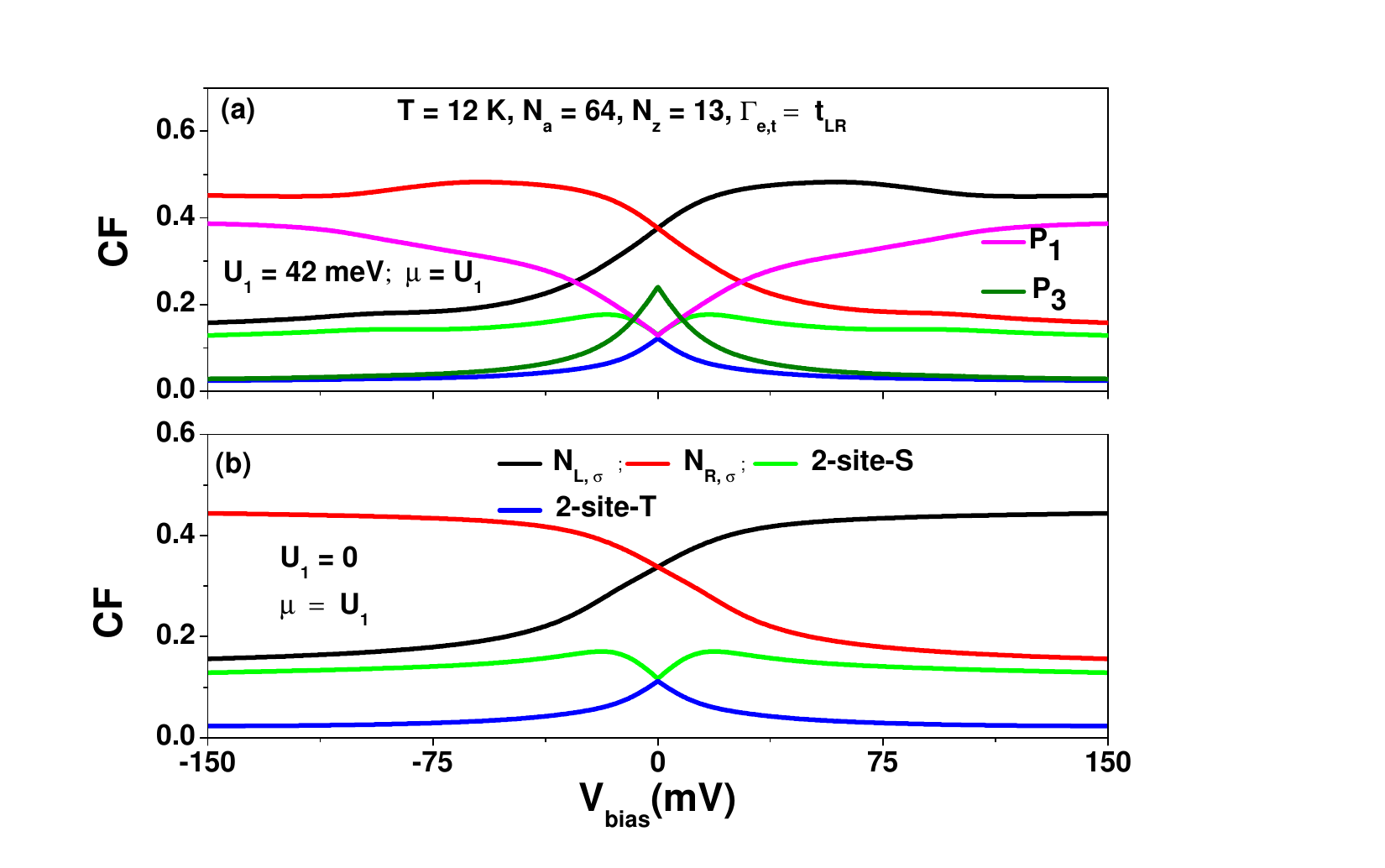}
\caption{The correlation functions ($CF$) of AGNRs with $N_a = 64$
and $N_z= 13$ at $E_L = E_R = 0$, $\mu = U_1$~meV, $\Gamma_{e,t}=
t_{LR}$ and $T = 12 K$, are plotted as a function of applied bias
in (a) $U_1 = 42$~meV and (b) $U_1 = 0 $. It should be noted that
$\mu_L = \mu + eV_{bias}/2$ and $\mu_R = \mu - eV_{bias}/2$ were
used in the calculation of the tunneling current. }
\end{figure}

To understand the current rectification behavior observed in Fig.
11, we present $N_{L(R),\sigma}$ and the inter-site two-particle
correlation functions (2-site-S and 2-site-T) for various
$\Gamma_{e,R}$ values at $\Gamma_{e,L} = 3$ meV in Fig. B.2. As
shown in Fig. B.2, the condition of $N_{L,\sigma}(V_{bias}) =
N_{R,\sigma}(-V_{bias})$ is lifted when the junctions have
asymmetrical contacted properties. $P_3$ channels dominate the
behavior of the tunneling current at low bias. Therefore, the
forward maximum current ($J_{F,1}$) and the reversed maximum
current ($J_{R,1}$) can be determined by
$P_3=N_{R,-\sigma}-\langle n_{R,-\sigma} n_{L,\sigma}\rangle$ and
$P_3=N_{L,-\sigma}-\langle n_{L,-\sigma} n_{R,\sigma}\rangle$,
respectively. In the low bias region, we have $N_{L,\sigma}\approx
N_{R,\sigma}$ and $\langle n_{R,-\sigma} n_{L,\sigma}\rangle >
\langle n_{L,-\sigma} n_{R,\sigma}\rangle$, which explains why
$J_{R,1}$ is larger than $J_{F,1}$ and the current rectification
behavior is observed in Fig. 11.

\begin{figure}[h]
\centering
\includegraphics[angle=0,scale=0.3]{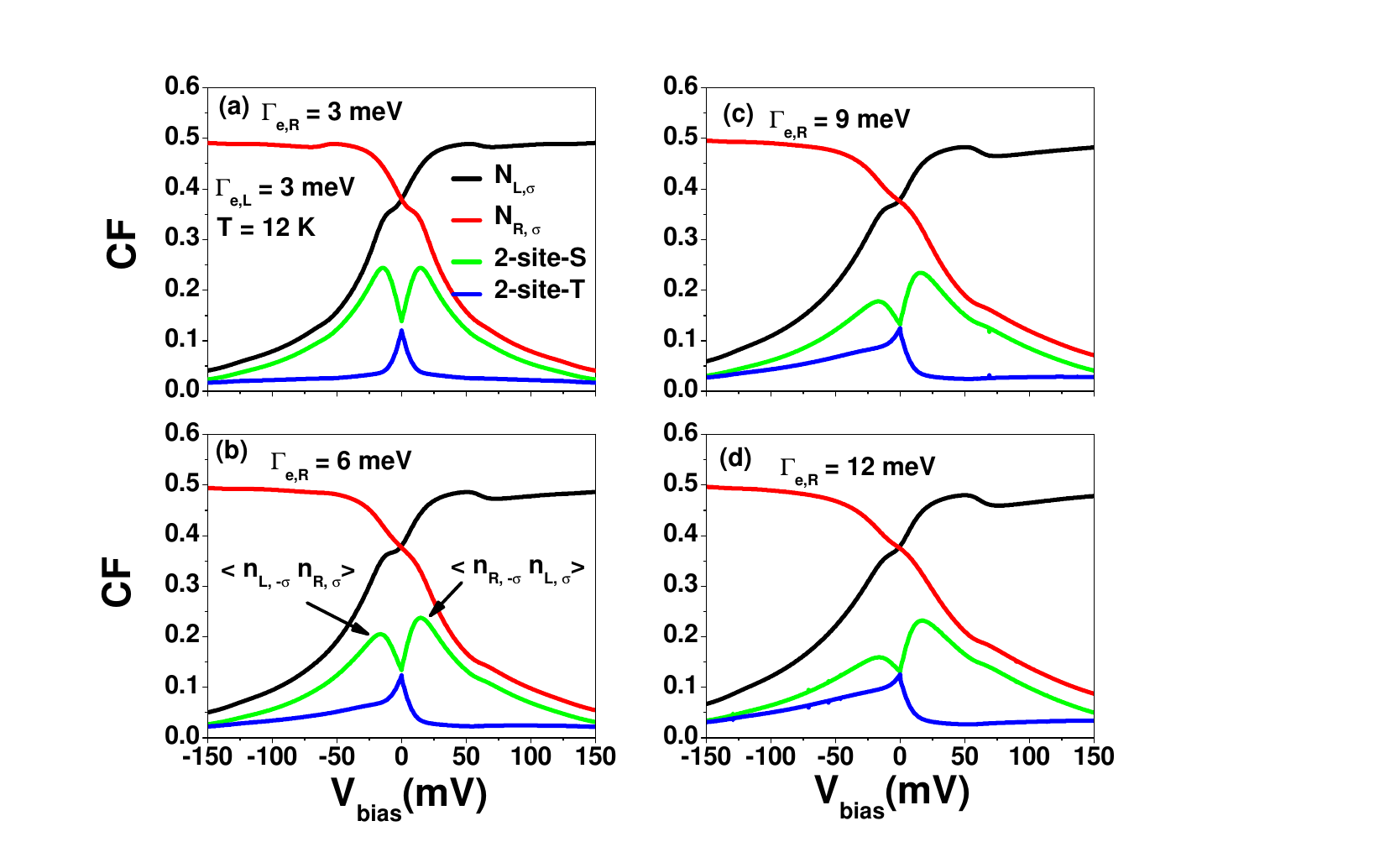}
\caption{Correlation functions $CF$ as a function of applied bias
at $E_L = E_R = 0$, $\mu = U_1 = 42$ meV, $\eta= 0.16$,
$\Gamma_{e,L}= 3$ meV and $T = 12 K$. (a) $\Gamma_{e,R}= 3$~meV,
(b) $\Gamma_{e,R}= 6$~meV, (c)$\Gamma_{e,R}= 9$~meV, and (d)
$\Gamma_{e,R}= 12$~meV. Other physical parameters are the same as
those of Fig. 11.}
\end{figure}




\setcounter{section}{0}
\setcounter{equation}{0} 

\mbox{}\\





\end{document}